\documentclass[%
 reprint,
 superscriptaddress,
 amsmath,amssymb,
 aps,
]{revtex4-2}

\usepackage{graphicx}
\usepackage{xfrac}
\usepackage{hyperref}
\hypersetup{colorlinks,urlcolor=blue, linkcolor=blue}
 
\begin{document}

\title{Surveying optically addressable spin qubits for quantum information and sensing technology}

\author{Calysta A. Tesiman}
\affiliation{Department of Materials and London Centre for Nanotechnology, Imperial College London, Prince Consort Road, London, SW7 2AZ, UK}

\author{Mark Oxborrow}
\email{m.oxborrow@imperial.ac.uk}
\affiliation{Department of Materials and London Centre for Nanotechnology, Imperial College London, Prince Consort Road, London, SW7 2AZ, UK}

\author{Max Attwood}
\email{m.attwood@imperial.ac.uk}
\affiliation{Department of Materials and London Centre for Nanotechnology, Imperial College London, Prince Consort Road, London, SW7 2AZ, UK}

\begin{abstract}
\section*{Abstract}
Quantum technologies offer ways to solve certain tasks more quickly, efficiently, and with greater precision than their classical counterparts. Yet substantial challenges remain in the construction of sufficiently error-free and scalable quantum platforms needed to unlock any real benefits to society. Acknowledging that this hardware can take vastly different forms, our review here focuses on materials that bear an optically-addressable electron or nuclear spin to embody qubits. Towards helping the reader to spot trends and pick winners, we have surveyed the various families of optically addressable spin qubits and attempted to benchmark and identify the most promising ones in each. We go on to reveal further trends that demonstrate how qubit lifetimes depend on the material's synthesis, the concentration/distribution of its embedded qubits, and the experimental conditions.
\end{abstract}

\maketitle
\section{Introduction}

Over the last 40 years, quantum computation has progressed from concept to hardware. In 2023, IBM reported the commissioning of a 1121-qubit superconducting quantum processor \cite{castelvecchiIBMReleasesFirstever2023}, while Honeywell demonstrated a quantum charge-coupled device architecture based on trapped ions \cite{pinoDemonstrationTrappedionQuantum2021}. Despite such achievements, the scalability and, thus ultimate utility of each species of hardware remains contested. 

Recently, materials that harbour an unpaired electronic or nuclear spin have garnered attention as contenders for use as qubit media. The best of these materials offer coherence times exceeding milliseconds, albeit at cryogenic temperatures \cite{Rancic2018}. Progress in the field has made the dream of usefully storing quantum information and/or implementing quantum processing at scale with appropriately interacting spins more tangible. For example, the start-up company Quantum Brilliance recently demonstrated a room temperature qubit system based on nitrogen-vacancy (NV) centres in diamond \cite{blackSupercomputersQuantumMachines2023}. Beyond computation, these same materials are already providing advantageous forms of quantum sensing, especially under ambient conditions where thermally-driven electronic and photonic noise present significant challenges for quantum applications \cite{barry2020}. Their advantages stem from the ability to optically initialise spin states into highly non-Boltzmann populations and, likewise, to implement optical read-out routines that are less impacted by blackbody radiation. 

\begin{figure}[b!]    
    \centering
    \includegraphics[width=\columnwidth]{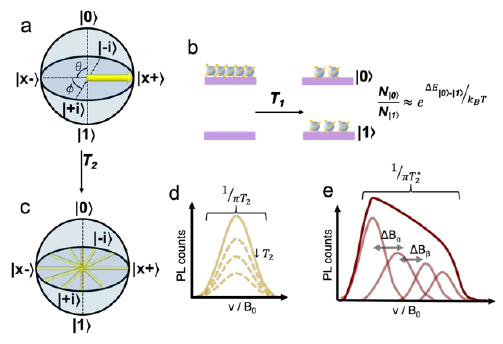}
    \caption{\textbf{Quantum spin parameters important for quantum technology.} (a) Bloch-sphere representation of a spin vector showing a quantum state composed of $|\psi\rangle = (|0\rangle + |1\rangle)/\sqrt{2}$. (b) Spin-lattice relaxation $T_1$ transforms an ensemble of polarised spins to a Boltzmann (thermal) spin distribution, approximated by the equation shown, where N$_{|0\rangle/|1\rangle}$ is the population of each state, $\Delta$E$_{|0\rangle-|1\rangle}$ is the energy difference between each state, $k_\textrm{B}$ is the Boltzmann constant, and $T$ is temperature. (c) Bloch sphere representing an incoherent collective superposition for an ensemble of spins after time $T_2$ has elapsed. (d) Homogeneous spin-resonance linewidth, demonstrating the impact of lowering $T_2$, and (e) the inhomogeneous linewidth broadened by neighbouring ($\alpha$ and $\beta$) magnetic spins; $\nu$ and $B_0$ represent the microwave frequency and magnetic field, respectively, as commonly used independent variables in quantum sensing.} 
    \label{fig:quantumparameters}
\end{figure}

In these quantum spin materials, information is most simply encoded in the spin state of a two-level system, but can be encoded across multiple levels in high-spin (S$\geq$1) systems; a property that may be used to facilitate certain error correction protocols \cite{Chizzini2022}. For qubits, a magnetic or electric field is applied to split otherwise degenerate spin states. The higher and lower energy spin states, $|0\rangle$ and $|1\rangle$, form two orthogonal basis states. A pure state is formed as a coherent superposition of the two ($|\psi\rangle = a|0\rangle + b|1\rangle$, see Figure \ref{fig:quantumparameters}a). Geometrically, we can visualise any such state as a point on the surface of a Bloch sphere, the location being defined by a polar angle $\theta$ and an azimuthal angle $\phi$ \cite{arecchiAtomicCoherentStates1972}. The value of $\theta$ determines the probability amplitude of finding the qubit in either the $|0\rangle$ or $|1\rangle$ state, while $\phi$ determines the relative phase of the qubit between the $|0\rangle$ and $|1\rangle$ states.

\begin{figure}
    \centering
    \includegraphics[width=1\columnwidth]{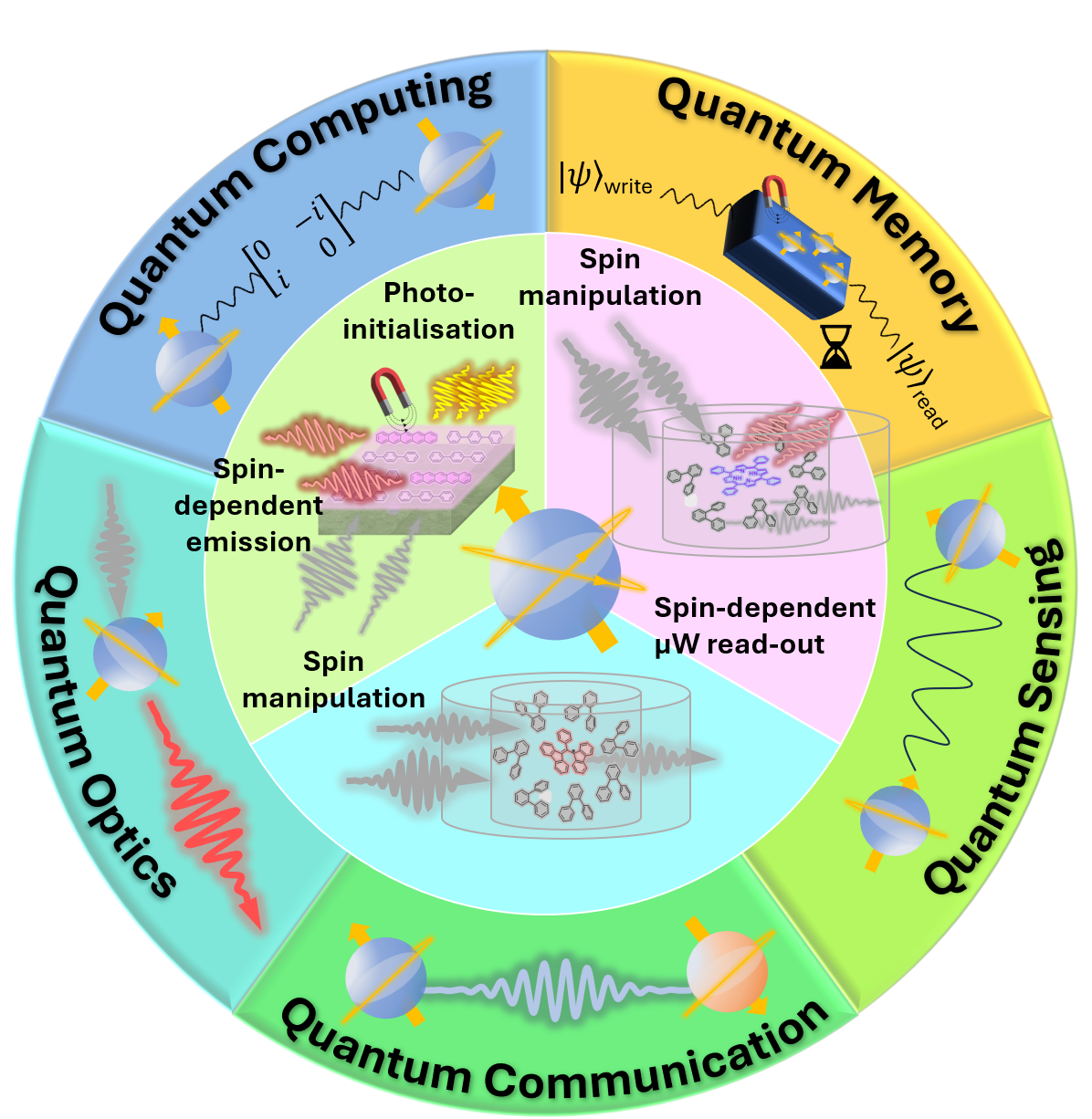}
    \caption{\textbf{Prominent approaches for addressing quantum spins for quantum applications (inner circle) and their applications in quantum technologies (outer circle)}. Optically-detected magnetic resonance (ODMR) spin measurements utilise spin-dependent luminescence with light/microwave-based spin manipulation (green third), whilst Electron Paramagnetic Resonance (EPR) spin measurements can employ light to initialise a spin system into a spin-polarised state (pink third) or one can employ microwaves to manipulate a thermally polarised system with spin-dependent microwave readout (blue third).}
    \label{fig:graphabs}
\end{figure}

To be useful, these materials must enable both reliable spin addressability and long storage times. These properties are, in turn, limited by the materials' spin-lattice relaxation ($T_1$) and spin decoherence times ($T_2$, also known as phase memory time). $T_1$ and $T_2$ represent the time it takes for the ``longitudinal'' and ``transverse'' magnitude of the spin state vector, respectively, to decay by a factor of e (Figure \ref{fig:quantumparameters}a $\&$ c). In practical terms, $T_1$ is the lifetime of a non-Boltzmann spin population (Figure \ref{fig:quantumparameters}b), and, put simply by DiVincenzo \cite{divincenzoQuantumComputersQuantum1999, DiVincenzo2000}, $T_2$ characterises the interactions of a spin qubit with its environment. These quantities are fundamentally related by $1/T_2 = 1/2T_1 + 1/T_\phi$, where 1/$T_\phi$ stems from contributions of so-called ``pure dephasing'' due to factors such as 1/f noise \cite{Bergli2009}, spin-bath fluctuations, and other magnetic/electric field gradients. Hence, in an idealised ``Markovian'' system, $T_2$ is only the decoherence time constant due to energy exchange with the environment (with the limit $1/T_2 = 1/2T_1$)\cite{Traficante1991}, and can be estimated by the homogeneous linewidth of a spin transition's resonant frequency ($\Delta_\textrm{hom} = 1/\pi T_2$, Figure \ref{fig:quantumparameters}d). In most cases, $T_\phi$ is significant and the transition linewidth becomes inhomogeneously broadened and can be used to estimate the overall spin dephasing, $1/T_2^* = 1/T_2 + 1/T_\phi = \Delta_\textrm{inhom}/\pi$, Figure \ref{fig:quantumparameters}e) \cite{Chavhan2009}. Sources of spin dephasing can arise from both intra- and intermolecular interactions and $T_2$ is limited by several controllable factors in the material's spin state, spin-orbit coupling (SOC), phonons, spin concentration, and spatial distribution of nuclear and electronic spins \cite{Chirolli2008}. Spin relaxation parameters are most commonly measured using electron paramagnetic resonance (EPR) techniques by measuring microwave absorptions and emissions from a material as a function of frequency or magnetic field strength, forming the basis for quantum spin technologies (Figure \ref{fig:graphabs}). 




Quantum systems capable of operating at room temperature open up many additional applications that the overhead of cryogenic operation precludes. But, achieving high fidelity (in the initialisation, gate operations and read-out) of room temperature qubits remains extremely challenging. In this review, we have attempted to survey the available ``fully-optically-addressable materials'' (FOAMs) that are most directly relevant to quantum information and sensing technology \cite{Roberts2025}. We do not provide an exhaustive account of all reported spin systems and their limitations. Rather, we present the different approaches, materials-wise, that have shown promise and report the properties (where available) of the best-performing representatives of each approach.

\section{Overview of Candidate Spin Qubit Materials}

\begin{figure*}
        \centering
        \includegraphics[width=\textwidth]{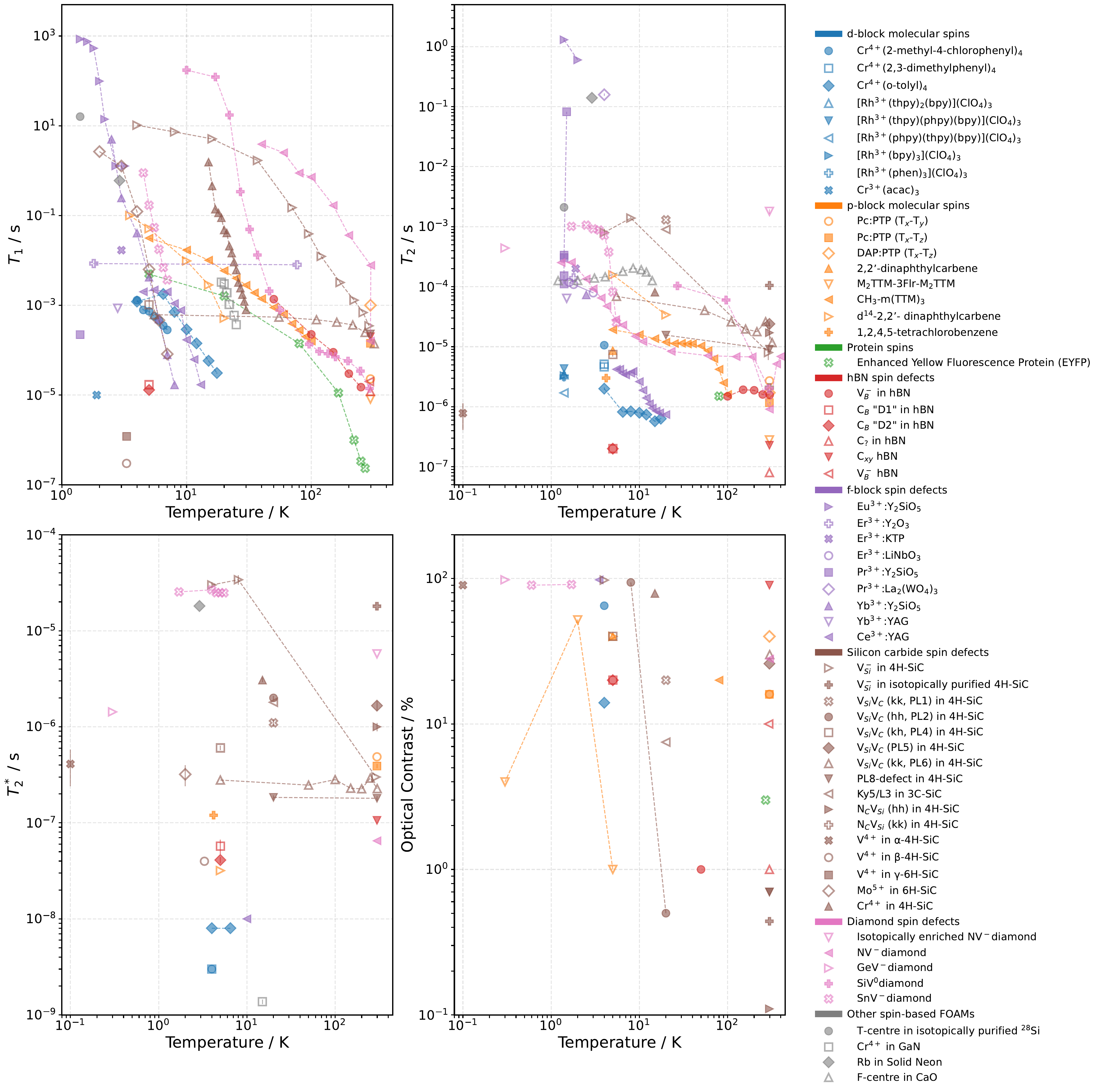}
        \caption{\textbf{A collection of selected quantum spin parameters measured using pulsed-ODMR techniques, including (a) spin-lattice relaxation, (b) spin coherence, (c) spin dephasing times, and (d) optical contrast.} Note: not every parameter has been reported for each system and conditions, hence not every material will be visible on all four plots. In each case, we have opted to include the highest measured parameter using standard inversion/saturation recovery, 3-pulse echo, Ramsey technique, regardless of appplied magnetic field. Data adapted from: 
        0.1$\%$ Pc:PTP \cite{Singh2025, Mena2024}, M$_2$TTM-3FIr-M$_2$TTM \cite{Chowdhury2024}, CH$_3$-m(TTM)$_2$ \cite{Kopp2025}, 2,2$^{\prime}$-dinaphthylcarbene \cite{Roggors2025}, 1,2,4,5-tetrachlorobenzene \cite{Breiland1975},     
        Cr$^{4+}$ molecular systems \cite{Bayliss2020, baylissEnhancingSpinCoherence2022}; Rh$^{3+}$ molecular systems \cite{Glasbeek2001},
        NV-diamond \cite{liuCoherentQuantumControl2019, takahashiQuenchingSpinDecoherence2008}, isotopically enriched NV-diamond \cite{Balasubramanian2009},
        SiV$^-$diamond \cite{Green2017, Sukachev2017}
        SnV$^-$diamond \cite{rosenthalMicrowaveSpinControl2023}, GeV$^-$diamond \cite{Senkalla2024};
        V$_\textrm{B}^-$ in hBN \cite{gottschollRoomTemperatureCoherent2021}, carbon defects in hBN \cite{Scholten2024, sternQuantumCoherentSpin, Chejanovsky2021}, 
        monovacancies in 4H-SiC \cite{Nagy2019,Babin2022}, isotopically purified SiC \cite{Lekavicius2022}, divacancies in 4H-SiC \cite{Christle2015, Crook2020, Li2022, Yan2020, Seo2016, linTemperatureDependenceDivacancy2021}, N$_\textrm{C}$V$_\textrm{Si}$ (hh) in 4H-SiC \cite{Wang2020b}, N$_\textrm{C}$V$_\textrm{Si}$ (kk) in 4H-SiC \cite{Jiang2023}, Cr$^{4+}$ in 4H-SiC \cite{koehlResonantOpticalSpectroscopy2017}, Mo$^{5+}$ in 6H-SiC \cite{Gilardoni2020}, V$^{4+}$ in 4/6H-SiC \cite{Wolfowicz2020}, Cr$^{4+}$ in GaN \cite{Koehl2017},   
        Eu$^{3+}$ in Y$_{2}$O$_{3}$ \cite{Zhong2019}, Er$^{3+}$ in Y$_{2}$O$_{3}$ \cite{bottgerOpticalDecoherenceSpectroscopy2024}, Y$_{2}$SiO$_{5}$\cite{Rancic2018}, KTP \cite{Bottger2016}, and LiNbO$_4$ \cite{Thiel2010, bottgerOpticalDecoherenceSpectroscopy2024}; Pr$^{3+}$ in Y$_{2}$SiO$_{5}$ \cite{equallHomogeneousBroadeningHyperfine1995} and La$_2$(WO$_4$)$_3$ \cite{Lovric2011}; Yb$^{3+}$ in Y$_{2}$SiO$_{5}$ \cite{Lim2018} and YAG \cite{Bottger2016b}; Ce$^{3+}$:YAG \cite{Azamat2017, Belykh2021}.
        Rb in solid Ne \cite{dargyteOpticalSpincoherenceProperties2021}, T-centre in $^{28}$Si \cite{Bergeron2020},
        EYFP protein \cite{Feder2024}, F-centre in CaO \cite{Glasbeek1981}.}
        \label{fig:AllCoherence}    
    \end{figure*}

FOAMs are an attractive option for quantum applications since they typically involve spin levels that are energetically separated by far more than $k_{\rm{B}} T$ and can be initialised using visible/near-infrared light to generate quasi-``pure'' quantum states. Furthermore, optical signals, composed of photons with energies far
larger than $k_{\rm{B}} T$, are less affected by thermal noise and benefit from the availability of single-photon sources and detectors capable of operating at high levels of fidelity even at room temperature.  FOAMS are often probed using ODMR spectroscopy, where the quantum sensing sensitivity for a.c. magnetic fields (for example), 
$\eta$, is proportional
to $\sqrt{t_{\textrm{overhead}}}/( C\sqrt{n_{\textrm{spin}} n_{\textrm{avgs}}} T_{2}^{*})$, where $t_{\textrm{overhead}}$ is the 
duration of the readout process, $C$ is the measurement contrast, $n_{\textrm{spin}}$ is the number of spins and 
$n_\textrm{avgs}$ is the number of scan averages. Relatively few FOAMs have been demonstrated so far and are often 
limited by either low photoluminescence yields or relatively short-lived spin polarisations 
($T_1$) and spin coherences ($T_2$), which have been summarised in Figure \ref{fig:AllCoherence} to support our discussion. As a prerequisite for inclusion in Figure \ref{fig:AllCoherence}, we have opted to follow the criteria put forward by Weber \textit{et al.,} for the realisation of an
optically addressable solid-state spin qubit (that is useable)\cite{Weber2010}:
    \begin{enumerate}
        \item A state must be paramagnetic and support two or more energy levels,
        \item An optical pumping cycle can be used to initialise the qubit,
        \item Luminescence to or from the qubit state varies by qubit sublevel in some differentiable way (i.e. intensity, wavelength, or other properties),
        \item Optical transitions must not interfere with the electronic state of the host,
        \item Differences between qubit sublevels must be large enough to avoid thermal excitation.
    \end{enumerate}

\subsection{3D spin-defect FOAMs}
\subsubsection{Diamond spin centres}

\begin{figure*}
        \centering
        \includegraphics[width=0.9\textwidth]{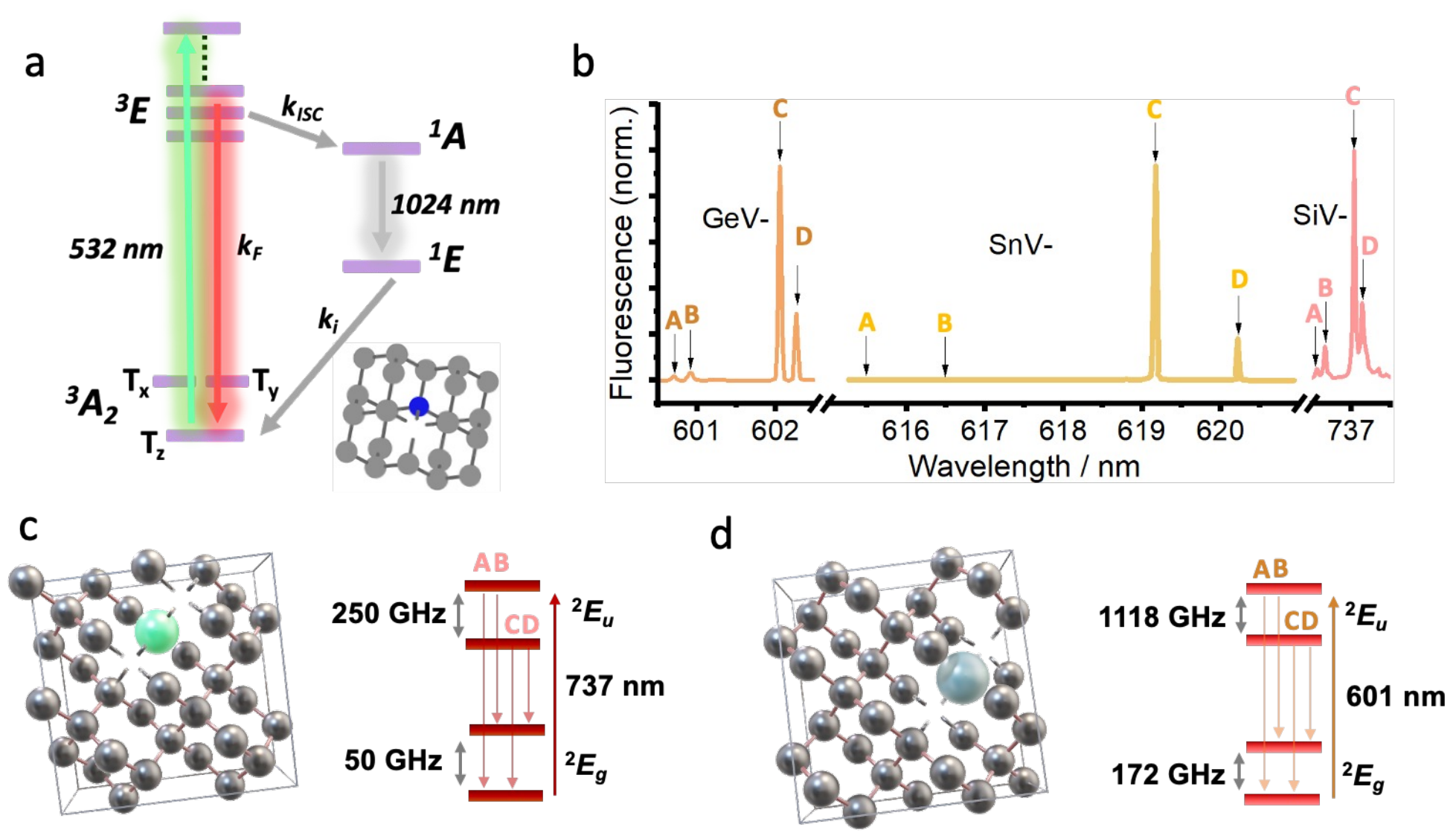}
        \caption{\textbf{Selected spin defects in diamond.} (a) Jablonski diagram for NV-diamond as a prototypical example of an optically addressable material; (b) low temperature fluorescence spectrum for SiV$^-$, SnV$^-$ and GeV$^-$ singlet spin centres showing electronic transitions between electronic and spin-orbit energy levels used for optically addressing spins; (c) example defect structure and energy level diagram for SiV$^-$ and (d) GeV$^-$ spin centres. Fluorescence data adapted from \cite{Karapatzakis2024, Trusheim2020, Muller2014}}
        \label{fig:diamond_defects}
    \end{figure*}

A well-established example of a FOAM which satisfies the above criteria is the negatively charged nitrogen-vacancy centres in diamond (herein simply, NV-diamond or NV$^-$-centres), which utilise a triplet ground state and the hyperfine splitting arising from the \textit{I}=1 $^{14}$N-nucleus. Initialisation (i.e., electron spin polarisation) is achieved by optical excitation with green light. The newly generated excited states either relax by emission of 637 nm light or undergo intersystem crossing (ISC) into a metastable singlet state (Figure \ref{fig:diamond_defects}a). Repopulation of the ground state then follows an intermediate relaxation between two singlet states (with emission at 1042 nm) and finally spin-selective ISC into the T$_0$ sublevel, resulting in a strong spin polarisation. At zero-applied magnetic field (ZF) the T$_{\pm1}$ states are degenerate due to the defect's $C_{3\textrm{v}}$ symmetry. The application of a magnetic field lifts this degeneracy, which enables spin manipulation using microwave pulses, thereby modifying the fluorescence at characteristic Zeeman splitting frequencies ($h \nu = g_\textrm{e} \mu_\textrm{B} B_0$), which can be detected following subsequent optical excitations and fluorescence back to T$_0$.

NV-diamond has been a cornerstone of ODMR-based quantum sensing due to its robust spin properties even at room temperature. Due to the relatively low spin densities (resulting in small dipolar coupling between spins) and the mismatch between diamond lattice vibrations (phonons) and the Larmor frequency of electron spins its S=1 ground state, NV-diamond can exhibit $T_1$s of several milliseconds at room temperature \cite{Mrozek2015, Cambria2023}. The nature and mechanism of its spin-lattice relaxation as a function of temperature, which includes an ``Orbach-like'' process (dependent on the phonon density at the spin transition frequency) and spin-phonon Raman scattering (of either first \cite{Norambuena2018} or else second \cite{Cambria2023} order) scaling like $T^5$, have been repeatedly investigated.

The material's quantum spin properties are highly dependent on the NV$^-$-centres's depth (below the diamond's surface) \cite{Zhang2017}, concentration \cite{Stanwix2010}, crystal strain \cite{Tsuji2024}, and the presence of impurities such as $^{13}$C, N-centres or EPR-inactive neutral or positively charged NV-centres \cite{Luo2022, Zhang2023}. Its popularity has seen a plethora of investigations to understand and modulate its spin properties. For example, high-field EPR spectroscopy and temperature-dependent measurements have demonstrated that decoherence from the $^{14}$N and $^{13}$C flip-flop fluctuations can be almost eliminated at low temperatures where the spin bath is polarised \cite{Takahashi2008}. Here, $T_2$ reached $\approx$ 250~$\mu$s at 2~K, following a sharp increase below 12~K in
high-temperature high-pressure (HTHP) diamond samples. Achieving similar properties under low-field and higher temperature conditions with thermally polarised nuclei is challenging and requires meticulous materials preparation and/or the use of dynamical decoupling methods. For example, the impact of parasitic nuclear spins and impurities was shown most remarkably by Balasuramanian \textit{et al.,} \cite{Balasubramanian2009}. Careful growth by chemical vapour deposition (CVD) on a diamond substrate using isotopically enriched feedstock led to just 0.3$\%$ $^{13}$C abundance and low levels of other paramagnetic impurities, resulting in $T_2 \approx $ 1.8~ms at room temperature. 

One limitation of NV$^-$-centres in diamond is their propensity to undergo charge-state conversion into a neutral and magnetically inactive state, NV$^0$, during high laser power excitation. This spin-independent process is reversible, but results in an EPR-inactive parasitic reservoir of NV$^0$ that results in the loss of spin polarisation by up to 31$\%$ during pulsed measurements \cite{Chen2015} and can even out-compete $T_1$ relaxometry \cite{cardosobarbosaImpactChargeConversion2023}.

More recently, increasing attention has been paid to using diamonds as a host for other spin defects owing to diamond's wide band gap, efficient thermal dissipation, and physical and chemical stability. As such, various magnetically-active colour centres have been investigated for quantum applications \cite{Thiering2020}. These include HV \cite{Komarovskikh2014}, BV \cite{Umeda2022}, OV \cite{Hartland2014}, so-called group-IV vacancy defects \cite{Bradac2019} such as SiV \cite{Pingault2017, CChen2024}, SnV \cite{Gorlitz2022, Karapatzakis2024}, GeV \cite{Iwasaki2015}, and PbV \cite{Trusheim2019,Wang2024}, and even transition metal defects originating as impurities like NiV \cite{Morris2024}. To our knowledge, spin polarisation has not been observed in HV or OV centres, whilst ODMR experiments on BV and PbV have not been reported. Despite this, PbV in particular, both in charged PbV$^-$ and neutral PbV$^0$ states is expected to exhibit robust spin coherence properties up to 9 K \cite{Mohseni2025}.   

Metal-vacancy spin-centres typically exhibit strong spin-orbit coupling (several hundred GHz) that gives rise to zero-field splitting (ZFS) even for S=1/2 species, as well as electron-nuclear spin coupling, making them qudit candidates (Figure \ref{fig:diamond_defects}b-d). Moreover, group-IV defects exhibit useful photonic properties that make them attractive for quantum applications such as Fourier-limited zero-phonon line (ZPL) linewidths, high spectral stability, coherent photon emission, and strain-responsive band gap engineering \cite{Meesala2018}. 

For example, negatively charged SiV (S=1/2 system, Figure \ref{fig:diamond_defects}c) has enabled direct observation of photon interference \cite{Waltrich2023}. Outside of a dilution refrigerator, the spin coherence lifetimes are severely impacted by thermal acoustic phonon coupling. At 100 mK, $T_1$ reaches 1 second while the longest $T_2^*$ measured was 1.5 $\mu$s. Spin coherence could be maintained by dynamical decoupling up to 600 mK where the coherence time during a dynamical decoupling scheme ($T_\textrm{DD}$) measured 60 $\mu$s. $T_2^*$ can also be improved through strain engineering, which modifies the spin-orbit coupling and is responsible for inducing the ground-state splitting, and subsequently the spin-phonon coupling \cite{Sohn2018}. By comparison, the neutral SiV (S=1 ground state) demonstrates an impressive $T_1$~$ \approx$ 25~s, $T_2$~$ \approx$ 0.1~ms even at 15~K, decreasing to 7.8 and 2~$\mu$s at room temperature \cite{Green2017}. However, a route to reliably synthesising SiVs in diamonds remains elusive. 

Negatively charged GeV (S=1/2 ground state, Figure \ref{fig:diamond_defects}d) has been investigated using ODMR for quantum memory applications at 300 mK and found to exhibit a $T_2$~$ \approx$ 440 $\mu$s and $T_2^*$~$\approx$ 1.46 $\mu$s \cite{Senkalla2024}. These defects were found to be particularly responsive to dynamical decoupling protocols with $T_\textrm{DD}$ reaching 24 ms, representing a significant improvement compared to negatively charged SiV.

SnV$^-$s (S=1/2 ground state) are also robust spin centres and have been the subject of spin control experiments. Rosenthal \textit{et al.,} report $T_1$ $\approx$ 20 ms and $T_2$ $\approx$ 170 $\mu$s when measured at 1.7 K in highly strained SnV-centres; these spin lifetimes afford  significant improvement in operation fidelity \cite{rosenthalMicrowaveSpinControl2023}.By comparison, Trusheim \textit{et al.,} report a longer $T_1$~$\approx$ 1.26 ms and $T_2^*$~$\approx$ 540~ns at 2.9~K in a less strained system \cite{Trusheim2020}.

Negatively charged NiV-centres (S=1/2) are near-infrared emitters, making them especially interesting for quantum communications due to their compatibility with conventional optical cables \cite{Thiering2020}. The S=1/2 ground state of the negatively charged NiV-centre has a predicted 0.1 ms coherence time at 4 K \cite{Thiering2021}, but only recently have steady-state ODMR studies been reported \cite{Morris2024}. 

\subsubsection{Silicon Carbide Spin centres}

\begin{figure*}[hbt]
        \centering
        \includegraphics[width=0.9\textwidth]{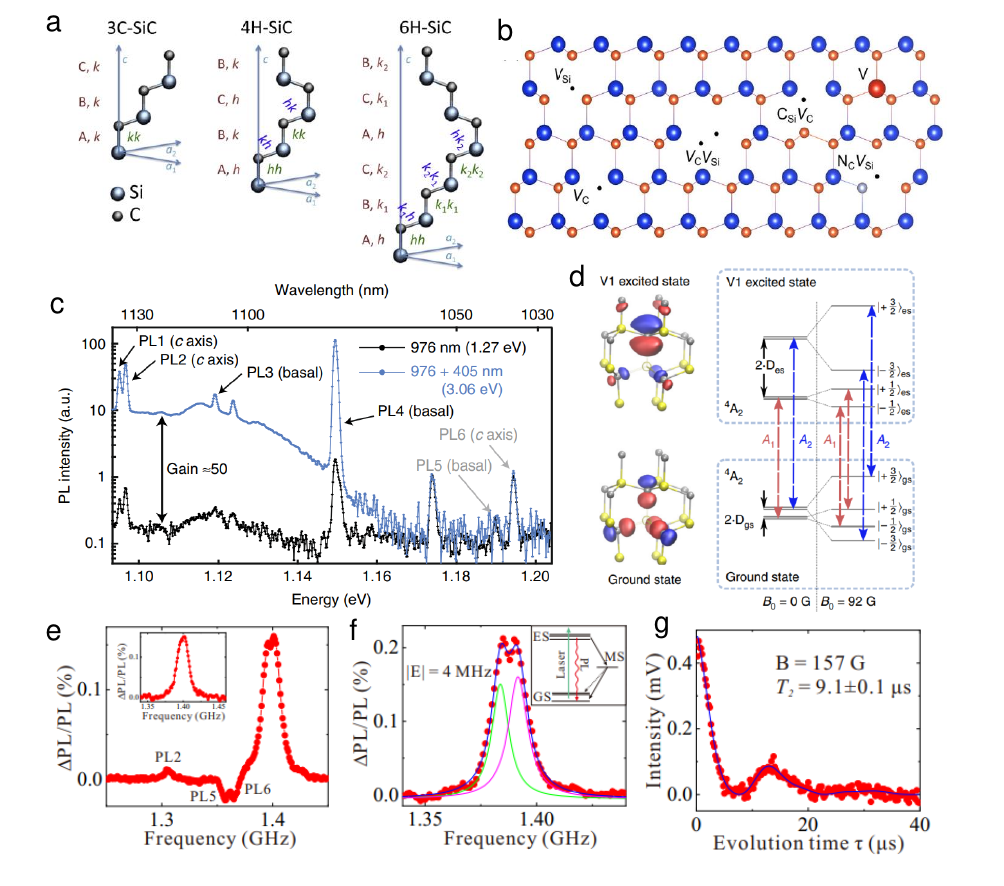}
        \caption{\textbf{Optically addressable point defects in SiC.} (a) Structures of the most commonly studied SiC polytypes, 3C-SiH, 4H-SiC and 6H-SiC showing the origin of divancancy nomenclature depending on the lattice substitution site; (b) 2D-depiction of selected SiC ODMR-active defects. (c) Photoluminescence spectrum from a SiC ensemble showing multiple individually addressable defect types (PL1-PL6). (d) Orbital distribution for the ground and excited states of the V$_B^-$-defect with the corresponding quartet spin-level energy diagram with and without an applied magnetic field. Red and blue arrows indicate optical transitions that connect m$_\textrm{s}$=$|1/2\rangle$ and m$_\textrm{s}$=$|3/2\rangle$ states, respectively. (e) ODMR spectrum of 4H-SiC samples at 20~K and (f) 300~K (inset optical initialisation scheme), demonstrating the robust spin-optical properties of the PL8-defect (the highest contrast peak), (g) the corresponding room-temperature Hahn-echo decay trace for PL8. Figures 1a, 1b, 1c, 1d, 1e-g were reproduced from references \cite{Davidsson2018, Bathen2021, Wolfowicz2017, Nagy2019, Yan2020}, respectively, with permission from MDPI and the American Chemical Society under the open access Creative Commons licence.}
        \label{fig:SiC_Figure}
    \end{figure*}

Beyond diamond, silicon carbide (SiC) also shows promising optical and coherence properties for quantum applications \cite{Fischer2018, Castelletto2022, Harmon2022, Gottscholl2022, Ou2024, Castelletto2024}. SiC is a complex material with more than 200 polymorphs. It is also used commercially in electronics and hence benefits from decades of manufacturing maturation. So far, the study of quantum systems has been largely restricted to 3C-, 4H-, and 6H-SiC, where C and H signify cubic and hexagonal structures, respectively, and the preceding number designates its polytype \cite{Castelletto2020} (see Figure \ref{fig:SiC_Figure}a). Pure SiC has a wide bandgap ($\approx$ 2 - 3~eV), weak spin-orbit coupling, and a naturally low abundance of nuclear spins. Importantly, it is capable of harbouring several varities of optically addressable colour centre with (often) near-infrared emission and record ODMR contrasts \cite{Nagy2019}. The most commonly studied defects encompass monovacancies (V$_\textrm{C}$ or V$_\textrm{Si}$) and divacancies but other defects may include carbon anti-site vacancies (C$_\textrm{Si}$V$_\textrm{C}$) (see Figure \ref{fig:SiC_Figure}b), charged NVs \cite{Wang2020b, Jiang2023}, Cr$^{4+}$- \cite{Koehl2017, Diler2020}, V$^{3+}$, V$^{4+}$ \cite{wolfowiczVanadiumSpinQubits2020, Ahn2024}, Mo$^{5+}$ \cite{Gilardoni2020}, and Ti-centres \cite{Baur1997, Castelletto2020, Castelletto2024}. Further layers of complexity are added by consideration of additional charge states and crystallographic sites within each polytype. Understandably, studies often focus on the ``best-performing'' defect within a given sample, and seldom are single samples of SiC homogeneous. This lack of homegeneity is perhaps the material's most significant practical limitation as a spin qubit candidate compared to other platforms.

Nevertheless, several defects exhibit outstanding spin-optical properties, and the fabrication challenges are beginning to be addressed. Negatively charged V$_\textrm{Si}$ in isotopically purified ($^{28}$Si) 4H-SiC exhibits the largest ODMR contrast at $\approx$ 97$\%$ at 4 K \cite{Nagy2019}. With an S=3/2 ground state, its associated ZFS is a few MHz with degenerate pairs of $|1/2\rangle$ and $|3/2\rangle$ states (Figure \ref{fig:SiC_Figure}d). Under a magnetic field ($\approx$ 82 mT) precisely aligned to the crystallographic c-axis, this degeneracy is lost and in the excited, so-called ``V1'' state, $|3/2\rangle$ shifts higher in energy than the $|1/2\rangle$ states. To achieve $\approx$ 97$\%$ contrast, the authors first equilibrate the spin populations using a 40 $\mu$s off-resonance pump (at 730 nm), followed by an on-resonance pump (at 861 nm) lasting up to 80 $\mu$s. On-resonance optical pumping causes $|3/2\rangle$ states to selectively decay into a non-radiative metastable state, followed by spin-selective repopulation of $|1/2\rangle$ ground states. This leads to a spin polarisation of up to 90$\%$ which is paired with record $T_2$s and $T_2^*$s of 0.8$\pm$0.12 ms and 30$\pm$2 $\mu$s, respectively. 

At room temperature, dynamic decoupling techniques can be used to extend $T_2$ from 8 $\mu$s to 47$\pm$20 ms \cite{Simin2017}. More recently, it was shown that V$_\textrm{Si}$ defects can be implanted into nanophotonic waveguides fabricated from 4H-SiC while also controlling the alignment of individual defects and maintaining excellent spin-optical properties \cite{Babin2022}. Here, $T_2^*$ of bulk V$_\textrm{Si}$-centres was measured at 34$\pm$4 $\mu$s at 10 K, whilst those in the $\approx$ 1 $\mu$m diameter waveguides measured at 9.4$\pm$0.7 $\mu$s. The coherence properties can be further improved by a factor of 10 through a regime of isotopic purification, and another factor of 5 by reducing strain inhomogeneity through a regime of annealing. Using this combined approach, Lekavicius \textit{et al.,} enhanced $T_2^*$ from 400 ns to $\approx$ 20 $\mu$s at room temperature \cite{Lekavicius2022}.

Divacancies also exhibit compelling spin-optical properties including up to 94$\%$ optical contrast \cite{Christle2015}, albeit at generally lower temperatures. Divacancies are formed by annealing pre-irradiated SiC at over 700$^o$C, however, the conversion efficiency into divacancies only ever reaches a few percent \cite{Falk2013, Sun2023b}, which may in part be due to counterproductive divacancy-dissociation back into V$_\textrm{C}$, V$_\textrm{Si}$, and potentially even antisite-C$_\textrm{Si}$V$_\textrm{C}$-centres under the right conditions \cite{Lee2021a}. Divacancies are most often given the generic label ``V$_\textrm{Si}$V$_\textrm{C}$'', designating that these neutral defects occur when both a carbon and silicon atom are missing from the lattice. However, it is important to note that depending on the polytype, there are several crystallographic sites available (Figure \ref{fig:SiC_Figure}a-b) and it is even possible to generate defects within stacking faults. Therefore, V$_\textrm{Si}$V$_\textrm{C}$-defects can exhibit either c-axis, $C_{3\textrm{v}}$, or basal-type, $C_{1\textrm{h}}$, symmetry. As a result, in 4H-SiC alone there are at least eight unique divacancies, the best known examples of which are sometimes colloquially called PL1-8, and are all distinguishable by their unique ZPL emission frequency (Figure \ref{fig:SiC_Figure}c) and are mostly assigned to have S=1 ground state due to their circa 1~GHz ZFS (Figure \ref{fig:SiC_Figure}e).

The hh-divacancy, PL2 in 4H-SiC, with natural isotopic abundance can demonstrate $T_2$s of 1.3 ms at 20~K when decoupled from $^{13}$C and $^{29}$Si nuclear spins under a 30 mT field \cite{Seo2016}; which is significantly longer than NV-diamond under similar conditions. Moreover, NV$^-$-centres in 4H-SiC (N$_\textrm{C}$V$_\textrm{Si}$) have known kk, hh, hk and kh-type defects and also exhibit robust spin parameters at room temperature \cite{Wang2020b, Jiang2023}. However, unlike their diamond analog, N$_\textrm{C}$V$_\textrm{Si}$s benefits from a NIR-emission with four distinct ZPLs between 1170 and 1250~nm; which is a preferred region for biosensing applications. Some of these divacancy centres also demonstrate remarkable insensitivity to temperature. For example, Yan \textit{et al.,} reported that despite having an unknown structure, PL8-centres exhibit similarly intense ODMR signatures at 1.4~GHz when measured at 20~K or room temperature \cite{Yan2020} (Figure \ref{fig:SiC_Figure}e-f). This temperature insensitivity is also reflected in the quantum spin properties where $T_2$ and $T_2^*$ measure 15.6$\pm$0.5 $\mu$s and 184$\pm$10~ns at 20~K, respectively, and 9.1$\pm$0.1~$\mu$s and 180$\pm$9~ns at room temperature, respectively (Figure \ref{fig:SiC_Figure}g). This long-lived room temperature spin coherence is shared to a lesser extent by PL6-centres, which are thought to be hh-divacancies occupying a stacking fault and are distinguishable by their 1038~nm (vs PL8's 1007~nm) ZPL emission. This demonstrates that several species in SiC are suitable for room temperature quantum applications. Moreover, divacancy systems were the first spin-centres demonstrated to be amenable to all-electrical spin-ensemble readout and initialisation schemes \cite{Klimov2014, Lew2024}. These functionalities have since been discovered with monovacancies under ambient conditions \cite{Niethammer2019}, demonstrating a strong potential to avoid some of the difficulties associated with ODMR spectroscopy, such as pump-light stability and photon collection efficiency.

Beyond vacancy systems, several metal-ion centres have demonstrated interesting spin-optical properties when implanted into SiC. The best to date in terms of its optical addressability is Cr$^{4+}$ in 4H-SiC. As a S=1 species, this material exhibits $T_1$~$>$1 second, with $T_2$ and $T_2^*$ = 81 $\mu$s and 317 ns, respectively, at 15 K \cite{Diler2020}. Importantly, it also demonstrates a 79$\%$ contrast, marking it as a system with one of the highest optical readout fidelities. The spin-optical properties of the same ion are markedly impaired in GaN host, which demonstrates 27$\times$ broader emission linewidths due to interactions of Cr$^{4+}$-centres with the surrounding spin bath\cite{Koehl2017}. 

V$^{4+}$ is perhaps the most thoroughly studied ion with coherent manipulation of its S=1/2 spin system reported in $\alpha$-4H-SiC \cite{Koller2025}, $\beta$-4H-SiC, and $\gamma$-6H-SiC \cite{Wolfowicz2020} host matrices. Whilst generally the spin relaxation parameters are shorter than Cr$^{4+}$, all V$^{4+}$ defects emit around 1.3 $\mu$m with markedly narrow inhomogeneous emission linewidth (ca.~100 MHz) and resolvable hyperfine coupling of the electron spin to the I=7/2 $^{51}$V nucleus, making them promising candidates for low-loss o-telecom band quantum applications \cite{Cilibrizzi2023}. Unlike most of the spin-defects discussed so far, spin polarisation of V$^{4+}$ is achieved using a magnetic field to lift the zero-field degeneracy of the $m_\textrm{s}\pm1/2$ states to generate two-distinct spin-dependent ZPLs which can be selectively depopulated using circularly polarised light \cite{Tissot2022}. This scheme has been used successfully to generate strong ensemble spin polarisation resulting in reported pulsed ODMR contrasts of up to 90$\%$ \cite{Astner2024}. Care is needed however, since use of an on-resonance (with the ZPL) initialisation scheme generates V$^{3+}$ states, which can be compensated for using an off-resonance green or UV repump protocol. 

Mo$^{5+}$ defects in 6H-SiC are another S=1/2 spin defect demonstrated to enable coherent spin manipulation at cryogenic temperatures. Here, rather than being a hindrance, the strong SOC of the heavy Mo$^{5+}$-ion is thought to protect it from parasitic phonon modes and spin-spin interactions, leading to $T_1$ of 2.7 seconds at 2 K \cite{Gilardoni2020}, significantly longer than V$^{4+}$. However, its ZPL is broader than vanadium at 24 GHz at 4 K, and shifted to 1120 nm \cite{Bosma2018}, which is slightly outside the highly desirable telecom-band.

\subsection{2D- and van der Waals FOAMs}

Van der Waals (vdW) materials are made of 2D-atomically thin layers weakly bound together through van der Waals interactions. These include graphene, hexagonal boron nitride (hBN) and transition metal dichalcogenides (TMDs). Despite a range of remarkable materials having been identified as ODMR-active, the precise structure and even spin multiplicity of the spin species is often ambiguous. Nevertheless, like other defect spin-qubit systems, inhomogeneous broadening and spin-phonon interactions are the principal causes of spin decoherence.

\begin{figure*}[hbtp]
        \centering
        \includegraphics[width=0.8\textwidth]{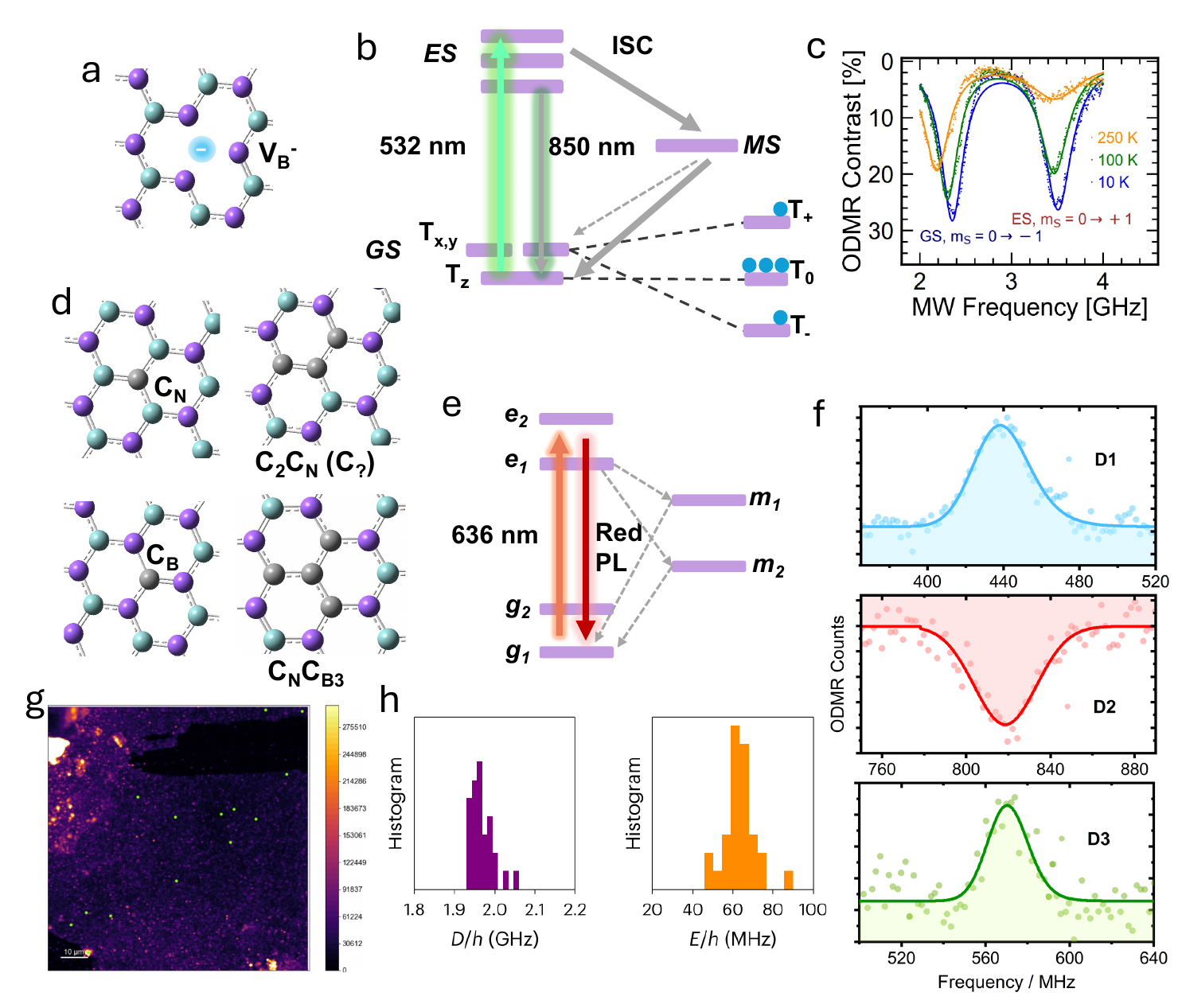}
        \caption{\textbf{Optically addressable spin defects in hBN.} (a) Proposed structure for the $V_\textrm{B}^-$ triplet spin centre (purple atoms = N, cyan atoms = boron and grey atoms = carbon, with (b) its corresponding Jablonski diagram and (c) variable temperature ODMR spectra at low field, confirming the defect's triplet character. (d) Proposed structures for selected carbon-based defects in hBN with (e) Jablonski diagram showing the involvement of two metastable states and repopulation of the g$_1$-state to achieve spin initialisation. (f) ODMR spectra for three carbon defects measured at 5 K showing g $\approx$ 2 doublet resonances corresponding to 154, 290 and 200 G for D1, D2, and D3-defects, respectively. (g) 2D-confocal map of hBN flakes showing fluorescent and ODMR-active (green dots) defect sites, with (h) histograms for 400+ carbon-based triplet defects. Figures (c), (f, data only), (g)-(h) were adapted from \cite{Mathur2022, Chejanovsky2021, sternQuantumCoherentSpin}, respectively, with permission from Springer Nature under the open access Creative Commons licence.}
        \label{fig:hBN_defects}
    \end{figure*}

The prototypical vdW material, graphene, was an early candidate for harbouring spin qubits; however, the metallic band structure of graphene is not conducive to forming luminescent spin centres. Instead, its isoelectronic analogue, hBN, has been the centre of remarkable advancement as a 2D-optically addressable platform at room temperature, building off its success as a single-photon source \cite{Shaik2021}. Magnetically-active spin centres in hBN were identified by EPR in the 1960s, with hyperfine structures giving clues to their identity as S=1/2 boron centres and carbon impurities \cite{Giest1964, Moore1972, Katzir1975}. However, the first optically addressable spin centres were only identified recently as triplets stemming from a negatively charged vacancy, likely originating from a missing boron atom, so-called $\textrm{V}_\textrm{B}^-$-centres \cite{Exarhos2019, Gottscholl2020, Mathur2022, Gong_2023} (Figure \ref{fig:hBN_defects}a). Similar to NV-diamond, optical initialisation is achieved by photoexcitation with 532 nm light, leading to ISC into a metastable singlet, which spin-selectively repopulates the fluorescent m$_\textrm{s}$=0 state (T$_\textrm{Z}$ at ZF, Figure \ref{fig:hBN_defects}b). The readout of the spin state occurs through 850 nm fluorescence with pulsed ODMR contrast of $\approx$ 1$\%$ at room temperature and almost $\approx$ 30$\%$ at 10~K, revealing a ZFS of $|D| \approx 3.5$~GHz (Figure \ref{fig:hBN_defects}c). Spin-relaxation and Hahn-echo measurements reveal a room temperature $T_1$ of 18 $\mu$s and $T_2$ of 2~$\mu$s \cite{gottschollRoomTemperatureCoherent2021}. These properties are particularly remarkable since both boron and nitrogen exhibit isotopically-abundant nuclear spins ($^{14}$N, I = 1, $^{10}$B, I = 3, and $^{11}$B, I = 3/2). In the same work, suppression of the nuclear spin bath using a second microwave source centred on a hyperfine line was found to improve $T_1$ and $T_2$ up to 25 $\mu$s and 7.5 $\mu$s, respectively. Work continues optimising the synthesis conditions for hBN to enhance the quantum properties of the defects. For instance, the contrast of V$_\textrm{B}^-$-centres can be increased to $\approx$ 10$\%$ at room temperature by controlling the material purity and electron irradiation dose, though $T_1$ appears to remain phonon limited \cite{healeyoptimisationElectronIrradiation2024}. 

Due to hBN's potential as a tunable single photon source, significant effort has been made to control substitutional-defect structures through careful synthesis. CVD-grown hBN has been found to contain a plethora of optically active carbon and oxygen-based spin species with stable ZFLs in the visible range \cite{Hayee2020, Kremarov2021}; although most have yet to be structurally characterised, or even the multiplicity conclusively determined (see Figure \ref{fig:hBN_defects}d for candidate structures and Figure \ref{fig:hBN_defects}e for the spin initialisation scheme). For example, three distinct so-called ``$\textrm{C}_\textrm{B}$''-defects, likely stemming from the substitution of a boron or nitrogen atom by carbon (``D1'', ``D2'', ``D3'', all likely S=1/2), demonstrate similar spin relaxation and coherence quantum properties ($T_2 \approx 0.2$~$\mu$s) with a continuous-wave (CW)-optical contrast up to 20$\%$ at 5 K \cite{Chejanovsky2021} (Figure \ref{fig:hBN_defects}f). Their inhomogeneous linewidths are also limited by the nuclear spin bath ($T_2^* \approx$~40-60~ns), where local spins also bestow an apparent ZFS of between 8-10 MHz. 

Though ensemble measurements remain challenging due to carbon-doped hBN's broad emission profile, a recent report has shown that it is possible to survey over 400 defect centres in CVD-grown samples. Collectively known as ``$\text{C}_\textrm{xy}$''-centres, careful measurements of individual spin-centres reveal distinctive triplet character with a ZFS of 15 MHz in contrast to most other accepted carbon-defects found in hBN \cite{Stern2022}. The triplet identity was later unambiguously characterised and confirmed for a series of defects with significantly larger ZFS with $|D|~\approx2$~GHz and $|E|~\approx 50-80$~MHz \cite{sternQuantumCoherentSpin} (Figure \ref{fig:hBN_defects}g and h). The $T_1$ at room temperature was measured to be 200$\pm$30 $\mu$s, $T_2\approx\,$1~$\mu$s and $T_2^*\approx\,$100~ns, competitive with V$_\textrm{B}^-$ defects. However, these properties are matched with a remarkable ODMR contrast of almost 50$\%$ (though it decreases with an applied magnetic field), giving rise to projected D.C. magnetic field sensitivity of up to 3 $\mu$T Hz$^{-1/2}$. A recent survey of similar defects demonstrated that under high microwave and optical pump powers, the optical contrast could reach as high as 90$\%$, though most exhibit $< 50\%$ despite similar ZFS, $T_1$ and $T_2$ \cite{Gilardoni2025}. The authors explore the potential of this material as a medium for D.C. vectorial quantum magnetometry by taking advantage of the defects' low symmetry and in-plane quantisation axis. Though not yet experimentally verified, the authors hypothesise that applying a field bias along a different material axis corresponding to each of the three triplet states, T$_\textrm{x}$, T$_\textrm{y}$, and T$_\textrm{z}$, should result in subtle changes in contrast that can be related to particular T$_n$ states to yield 3-vector components of the applied field. 

More complex carbon species demonstrating similar coherent behaviour at room temperature have been explored using metal-organic vapour-phase epitaxy with carbon-based additives \cite{Guo2023, Scholten2024}. The so-called ``$\textrm{C}_\textrm{?}$''-defect (potential identity C$_2$C$_\textrm{N}$) is another potential S=1/2 species or, two weakly coupled S=1/2 centres, has recently been reported as being co-located with V$_\textrm{B}^-$-centres following electron irradiation of hBN flakes \cite{Scholten2024}. While their $T_1$ and $T_2$ are similar to C$_\textrm{B}$-species ($\approx$ 13 $\mu$s and $\approx$ 80 ns, respectively), it can be used to ``sense'' V$_\textrm{B}^-$-centres through spin cross relaxation at an appropriate applied magnetic field. Due to its out-of-plane magnetic quantisation axis (versus the in-plane axis of V$_\textrm{B}^-$-centres), it can also be used more effectively to image magnetic structure on underlying substrates with an estimated DC magnetic sensitivity of $\approx$ 1 $\mu$T Hz$^{-1/2}$. 

One limitation of hBN spin defects is their propensity to form dark ``trap'' states that lead to blinking luminescence \cite{Martnez2016, Stern2019}. Unlike NV-diamond, the formation of these dark trap states is not dependent on laser power, and whilst the origin of this behaviour is still largely unknown, defects in multilayer hBN which are more protected from the environment are less prone to blinking \cite{Tran2016}. This suggests that atmospheric control may be a sensible approach to improve the reliability of spin-dependent luminescent read-out.   

\subsection{Molecular FOAMs}
\subsubsection{p-block molecular spin centres}

Molecular systems are becoming increasingly popular and offer an enticing opportunity to develop chemically tunable quantum materials catered to different applications \cite{Fataftah2018, Atzori2019, Wasielewski2020, Yu2021}. Ground-up synthesis enables the incorporation of particular functionalities such as stable radicals \cite{Wasielewski2023, Quintes2023}, modulation of triplet/singlet yields, enrichment with low or zero nuclear magnetic moments such as deuterium, oxygen and sulfur, or the targeted inclusion of nuclear spin-active elements such as nitrogen, phosphorous, transition metals, and lanthanides (\textit{vide infra}). A synthetic approach also enables changes to the host matrix \cite{baylissEnhancingSpinCoherence2022}, spin concentration, defect orientation, and material processing approaches, which are limited for defect-based systems. 

\begin{figure*}
        \centering
        \includegraphics[width=1\textwidth]{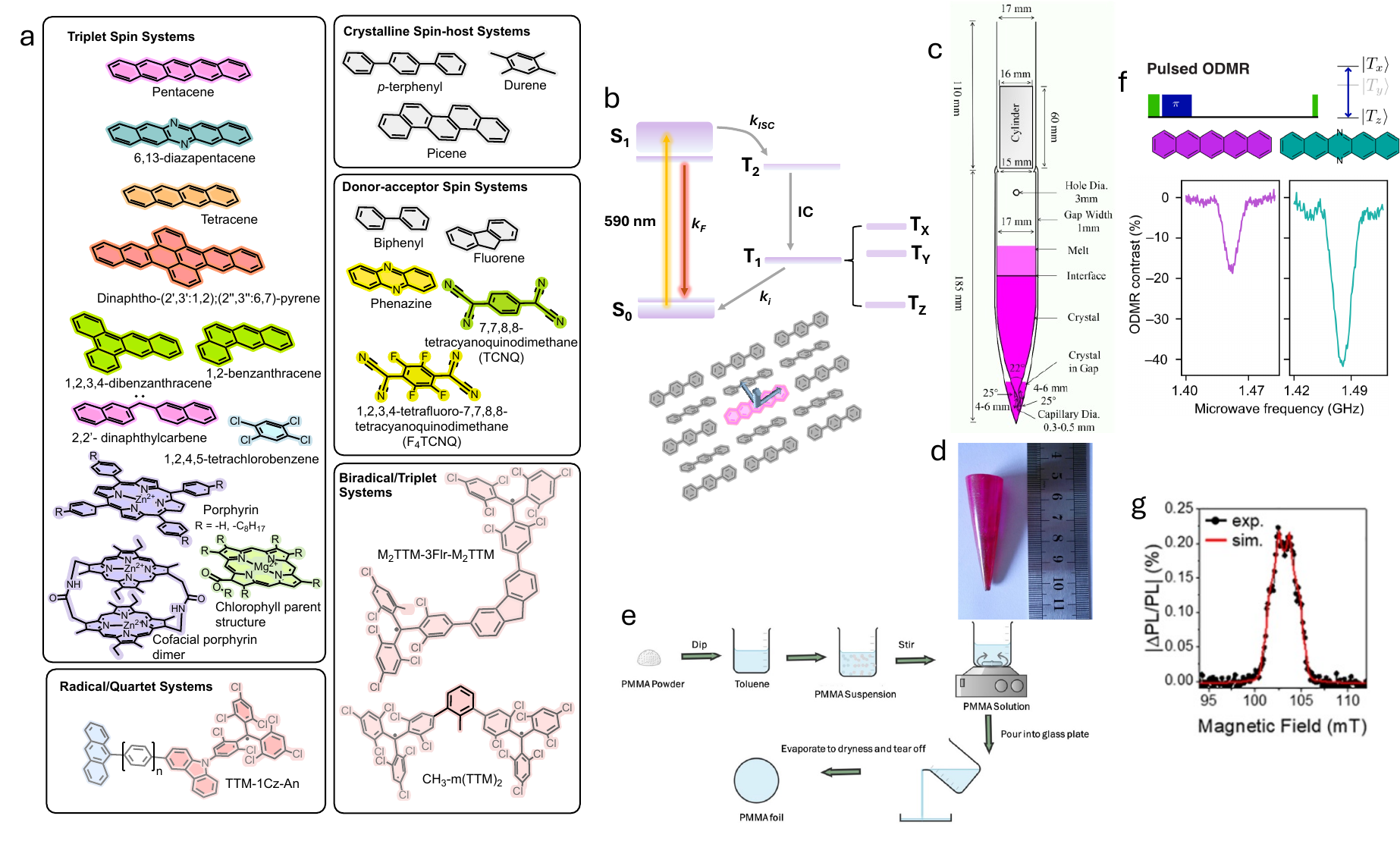}
        \caption{\textbf{Fully optically addressable organic quantum materials.} (a) Molecular structures of known organic FOAMs. (a) Jablonski diagram for ground-singlet state materials, like Pc:PTP, with inset doped structure crystal of p-terphenyl. The materials are excited/initialised using light. The spin centres subsequently relax by fluorescence ($k_\textrm{F}$) or populating the triplet states through intersystem crossing ($k_\textrm{ISC}$). Repopulation of the singlet ground state occurs in a triplet sub-level dependent manner ($k_i$, where \textit{i} is the corresponding triplet sub-level) and gives rise to high-contrast spin-optical signals. (c) Bridgmann growth set up to manufacture doped organic crystals such as (d) single crystals of Pc:PTP. (e) ODMR spectrum of Pc:PTP and DAP:PTP demonstrating enhanced pulsed optical contrast due to modified spin dynamics. (f) Synthesis methodology to generate optically clear films (aka ``foils'') of doped PMMA as an excellent amorphous host for radicals with spin-optical readout, with (g) example ODMR spectrum of CH$_3$-m(TMM)$_3$ in PMMA. Figures 1c-1d, 1e, 1f and 1g were reproduced from references \cite{Cui2020, Samal2023, Mann2025, Kopp2025} with permission from MDPI and the American Chemical Society under the open access Creative Commons licence.}
        \label{fig:pblock_FOAMs}
    \end{figure*}

The first examples of ODMR performed on molecular systems focused on small organic molecules \cite{Breiland1975, Hirota1979} and biologically-relevant porphyrin molecules such as those found in chlorophyll and bacterial reaction centres. These molecules use photoexcited triplet states as the magnetically-active spin centres that spin-selectively relax through phosphorescence, or relax to the fluorescent single states  \cite{Clarke1982, Kamyshny1992} (Figure \ref{fig:pblock_FOAMs}a-b). Porphyrins may be coordinated with Mg$^{2+}$ or Zn$^{2+}$ ions, which, though affecting the yield of the photoexcited spin state, do not themselves directly harbour it due to their filled s- or d-orbitals. Until recently, studies on p-block FOAMs were typically limited to cw measurements at cryogenic temperatures and aimed at investigating the interplay between ground and excited states, the influence of stereochemistry of the porphyrin dimer molecules \cite{Guckel1984}, and detecting triplet states in nanolayers \cite{Schaafsma2005}. A few groups did however extend their work to pulsed measurements, including Breiland et al.\cite{Breiland1975}, who measured the coherence properties of 1,2,4,5-tetrachlorobenzene in durene at ~4 K, and found a $T_2$ of a few microseconds, which could be significantly extended to a millisecond or so using dynamic decoupling or spin-locking techniques.

Single-molecule ODMR spectroscopy of pentacene molecules in a para-terphenyl matrix (Pc:PTP) has already been demonstrated at cryogenic temperatures in a series of remarkable works by Wrachtrup and colleagues \cite{Wrachtrup1993, Wrachtrup1993b, Kilin1998}. Only recently have pulsed experiments been performed to reveal contrast and spin coherence properties that are competitive with materials like NV$-$diamond at room temperature. The T$_\textrm{x}$-T$_\textrm{y}$ spin transition of a 0.1$\%$ crystal of Pc:PTP, grown using the Bridgman technique (Figure \ref{fig:pblock_FOAMs}c-d) demonstrates
$T_1 \approx \,23$~$\mu$s, $T_2 \,\approx \,2.7$~$\mu$s and $T_2^* \, \approx\,$500~ns\cite{Singh2025, Singh2024b}. Investigations using the more strongly spin polarised T$_\textrm{x}$-T$_\textrm{z}$ transition in both crystals and 100~nm-thin films at 0.01$\%$ and 0.1$\%$, respectively, reveal similar spin dynamics and also suggest an ability to modulate Pc:PTP's spin properties according to sample thickness and spin concentration\cite{Mena2024}. These robust spin-optical properties can be maintained in Pc:PTP nanocrystals down to 200~nm with an optical contrast similar to equivalently sized NV-diamond nanoparticles \cite{Ishiwata2025}. When coated with a Pluronic F-127 polymer, these nanoparticles also display excellent biocompatibility with low toxicity over 72 hours according to \textit{in vitro} explorations, paving the way for their use in cell assay techniques. 

The potential of using synthetic chemistry to modulate the spin dynamics to improve spin-optical properties has also been explored (see Figure \ref{fig:pblock_FOAMs}b). Substituting two carbon atoms on pentacene for nitrogen atoms resulted in 6,13-diazapentacene, which could be doped into a para-terphenyl host (DAP:PTP) by crystal growth and vapour deposition techniques \cite{ngMoveAsidePentacene2022, Mann2025}. While maintaining a relatively long $T_2$ of 1.46~$\mu$s and strong triplet spin polarisation, DAP:PTP exhibited an enhanced contrast of up to 40$\%$, surpassing the state of the art in NV-diamond platforms \cite{Mann2025} (Figure \ref{fig:pblock_FOAMs}e). This improvement was attributed to the introduction of new low-energy non-bonding states that enhance vibrationally-induced spin-orbit coupling between the ground-state singlet and the metastable T$_\textrm{x}$ state. It was also shown that direct solvent techniques can be used to grow ODMR-active DAP:PTP nanoparticles, which could be compatible with \textit{in vitro} sensing experiments.

Further organic systems have demonstrated potential as optically addressable materials, though, to our knowledge, pulsed optically detected experiments have yet to be performed. For example, the room temperature steady-state ODMR contrast of a 1$\%$ crystal of pentacene-doped picene has been measured at 15$\%$, representing a potentially significant improvement over the PTP matrix \cite{Moro2022}. Steady-state ODMR signals have also previously been reported at 2 K for (perdeutero)tetracene, 1,2-benzathracene, 1,2,3,4-dibenzanthracene \cite{Clarke1976} and dinaphtho-(2',3':1,2);(2",3":6,7)-pyrene \cite{Brauchle1979}. Interestingly, work by Corvaja, Pasimeni and Giometti \textit{et al.,} have shown that even highly spin-dense charge-transfer (CT) co-crystals can exhibit bright room temperature ODMR signals. Co-crystals comprised of donors such as biphenyl, fluorene, phenazine and acceptors such as 7,7':8,8'-tetracyanoquinodimethane (TCNQ) and 1,2,3,4-tetrafluoro-TCNQ (F$_{4}$TCNQ) have been studied to elucidate their triplet state dynamics \cite{Agostini1993, Gundel1989}. These materials exhibit narrow resonance lines due to intermolecular site hopping of triplet excitons. The resonances for each site can become resolved at low temperatures where hopping is not thermodynamically favoured \cite{Agostini1993}. Their high spin densities ($\approx \, $50$\%$) are highly advantageous for quantum sensing, where the a.c. sensitivity is proportional to the $\sqrt{n_\textrm{spin}}$, though this is often concurrent with less robust quantum spin properties compared with dilute materials such as Pc:PTP. Moreover, these materials are also promising hosts for the study of exotic spin behaviours like singlet fission and triplet-triplet annihilation \cite{Corvaja2005}. For example, at room temperature, the triplet states of phenazine:TCNQ are predominantly formed by singlet fission and using transient nutation EPR spectroscopy the authors estimate $T_1$ and $T_2$ times of $\approx$ 1 $\mu$s and 600 ns, respectively, even at room temperature \cite{Corvaja1992}. 

Several ground-state radical and diradical materials with an optical readout capacity have also been demonstrated \cite{gorgonReversibleSpinopticalInterface2023a, Chowdhury2024, Kopp2025}, as have materials with quintet states \cite{Joshi2022, Sun2023, Grune2024}. Diradical materials have the advantage of exhibiting a ground-state triplet, making their optical addressability protocol identical to NV-diamond. Optical readout is enabled by employing luminescent radical moieties based on tris(2,4,6-trichlorophenyl)methyl (TTM) \cite{Zhu2025}, where the radical is sterically isolated on a carbon atom with only weak hyperfine coupling to neighbouring $^{13}$C and hydrogen atoms. As a result, these materials can benefit from relatively narrow resonance linewidths, long $T_1$s, and can be embedded in frozen solutions or optically transparent PMMA foils through simple solvent-based methods \cite{Samal2023} (Figure \ref{fig:pblock_FOAMs}f-g).   

Additionally, ODMR has long been used to investigate genuine bio-molecules such as nucleotides and proteins using phoshphorescence of pyridine dinucleotides or certain amino acid residues such as tryptophan \cite{Kwiram1982, Clarke1982}, albeit at cryogenic temperatures. The emission wavelength-dependence of tryptophan's ZFS parameters enabled investigators to distinguish between different proteins and even different tryptophan residues in the same protein molecule. The usefulness of these studies was restricted due to the perceived inability of these molecules to maintain long $T_1$s at room temperature, where dynamics effects occur. However, recently fluorescent proteins have been demonstrated as a viable spin qubit medium \textit{in vivo} and in solution at room temperature. Feder \textit{et al.,} used ODMR spectroscopy to demonstrate that at 80 K the X-Z and Y-Z triplet transitions of enhanced yellow fluorescent protein (EYFP) demonstrated a 44$\%$ and 32$\%$ ODMR contrast, respectively \cite{Feder2024}. This corresponds with a zero-applied field $T_1$ of 141 $\mu$s (estimated from spin polarisation decay) and $T_2$ of 1.5~$\mu$s. Using Carr-Purcell-Meiboom-Gill (CPMG) dynamical decoupling the effective decoherence time ($T_\textrm{DD}$) reached 16~$\mu$s. The authors largely circumvent overhead limitations from long triplet lifetimes (ms) using an additional near-infrared pulse to induce T$_1$ to T$_2$ transitions to induce reverse ISC and delayed fluorescence from S$_1$. This interesting methodology could be suitable for other spin systems with quasi-resonant S$_1$ and T$_2$ electronic states, such as pentacene, to increase their sensitivity by increasing measurement repetition rates.

\subsubsection{d-block molecular spin centres}

Optically addressable d-block molecules have been investigated for several decades, though due to the weak oscillator strength of forbidden d-d transitions, the majority of ODMR studies focused on luminescence stemming from ligand-based triplet states generated following intramolecular ligand-to-ligand (LLCT) or metal-to-ligand charge transfer (MLCT) events \cite{Braun1989, Kamyshny1992}. These studies have generally focused on the use of diamagnetic heavy metal centres such as Pd$^{2+}$, Gd$^{2+}$, Rh$^{2+}$, Au$^+$ \cite{Weissbart1994} where the extent the metal-ligand orbital mixing is reduced. However, for elements with partially filled d-orbitals, the strong SOC stemming from the heavy metal centres plays a significant role in determining the spin-polarisation, spin-relaxation, and ZFS parameters. Accordingly, cryogenic temperatures are required to maintain the detectable transitions due to strong SOC effects of the metal centres. 
One of the first attempts to investigate the quantum spin properties of these materials focused on Rh$^{3+}$ biphenyl-type chelating ligands, where $T_\textrm{m}$ was measured at a few microseconds at 1.4~K\cite{Westra1991, Glasbeek2001}. Due to the deeply cryogenic conditions $T_1$-type processes were assumed to be negligible.

    \begin{figure*}[hbtp]
        \centering
        \includegraphics[width=0.8\textwidth]{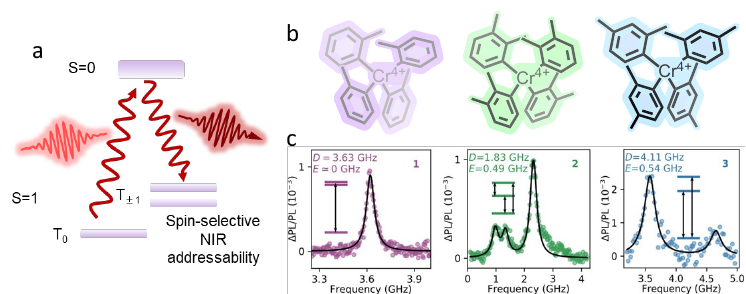}
        \caption{\textbf{Optically addressable Cr$^{4+}$ molecular colour centres.} (a) Energy level illustration of spin initialisation and readout using resonant NIR-photoinduced ISC from the ground state triplet to an excited singlet state, followed by spin-selective fluorescence to $T_\pm$ spin sublevels. (b) Molecular structures of three Cr$^{4+}$ tolyl-type complexes, and (c) their corresponding cw-ODMR signals. (a) and (c) were adapted from ref \cite{Bayliss2020}.}
        \label{fig:dblock_FOAMs}
    \end{figure*}

In the early 2000s, further ODMR-studies of transition metal complexes became far more scarce before lighter metal complexes regained attention in the 2010s. However, optically addressable materials with first row transition metal-spin centres were first reported using tris(aceylacetonato)Cr$^{3+}$ (acac) supported in an Al(acac)$_3$ crystalline host matrix at 1.9 K. While the ODMR properties of the S=3/2 ground state are not reported, on-resonance excitation to the lowest energy doublet state is sufficient to generate spin polarisation, with the ODMR signal detected through the vibronic side bands \cite{Fields1986}, making it a molecular analog of earlier ruby ODMR experiments \cite{Geschwind1965}. 

Optically addressable d-block molecules with a ground-state and zero-applied magnetic field spin manifold have only recently been demonstrated with Cr$^{4+}$ (Figure \ref{fig:dblock_FOAMs}a). The Cr$^{4+}$ spin-centre was realised with several tolyl-based ligand systems where modifications to the ligand structure and corresponding ligand field strength can be used to control optical excitation frequency and the ZFS of the Cr$^{4+}$ S=1 ground state \cite{Laorenza2021} (Figure \ref{fig:dblock_FOAMs}b-c). To reduce intermolecular dipole coupling, the spin-active moiety is diluted in a matrix with the S=0 isostructural tin analogue in crystals grown from hexane solutions \cite{Bayliss2020}. Optical spin polarisation is then achieved by exciting molecular spins from the S=1 state to the S=0 manifold, which fluorescently decays within a few microseconds to favourably populate the T$_\pm$ states. Initial pulsed ODMR experiments were performed on Cr$^{4+}$(o-tolyl)$_4$ spins, which benefit from an E=0 triplet and a relatively long $T_1$ of 0.22 ms and $T_2$ of 640 ns at 5 K. This permits several optical cycles to build spin polarisation and enhance the optical contrast up to 14$\%$. The coherent properties can be significantly improved using a non-isostructural Sn(4-fluoro-2-methylphenyl)$_4$ host matrix \cite{Bayliss2022}. Here, the ZFS is significantly increased, giving rise to clock transitions similar to those realised by the so-called ``zero first-order Zeeman'' (ZEFOZ) method, initially pioneered for lanthanide spin systems (see below). As a result, $T_1$ and $T_2$ are increased to $\approx$ 1.21$\pm$0.02 ms and $\approx$ 10.6$\pm$0.2 $\mu$s at 5 K, respectively, while the pulsed contrast can be improved up to 65$\%$. 

Work continues to explore alternative d-block elements where strong ligand-field interactions can enable optical excitation between spin manifolds and spin-dependent fluorescence \cite{Fataftah2018, Yu2021}. Candidates include Ni$^{2+}$ \cite{Wojnar2020, Wojnar2024}, Mo$^{4+}$ \cite{Laorenza2024}, V$^{3+}$ \cite{Fataftah2020, Laorenza2024}. Current challenges involve tuning the energy of the intermediate (e.g., singlet) state to provide the right conditions for spin polarisation, and identifying the appropriate ligand design to optimise photoluminescence.

\subsubsection{f-block molecular spin centres}

Lastly, there has been significant interest in f-block molecular systems. Here, we make a distinction between molecular systems where the ion is doped into a molecular lattice, and trapped ion systems where a single ion is levitated using electrostatic interactions to give rise to extremely long coherence times, but are not subject to qubit-quality improvements through molecular engineering \cite{Bruzewicz2019, Brown2021, Moses2023}. 

\begin{figure*}
        \centering
        \includegraphics[width=0.7\textwidth]{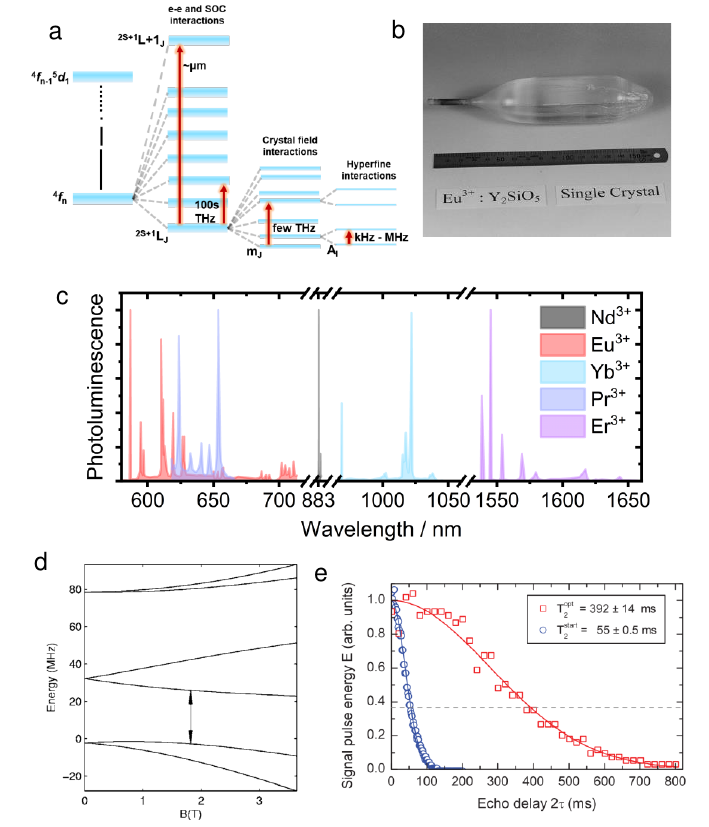}
        \caption{\textbf{Optically addressable lanthanide spin qubits.} (a) Generic energy level diagram showing energy level splitting due to electron-electron interactions (e-e, leading to $^{2S+1}$L states), spin-orbit coupling (SOC, \textit{J}), crystal field (m$_j$) and hyperfine interactions (A$_\textrm{I}$). (b) Example single crystal of Eu$^{3+}$:Y$_2$SiO$_4$ grown by the Czochralski method. (c) Collated example photoluminescence spectra from purely inorganic lanthanide-doped crystal host systems demonstrating coverage of the visible and near-infrared spectrum. (d) Examples of a zero first-order Zeeman (ZEFOZ) transition for Eu$^{3+}$:Y$_2$SiO$_5$ spin system with the transition region indicated, demonstrating equal energy-field gradients, resulting in magnetic field insensitivity, and (e) impact of the ZEFOZ method on spin decoherence time for Pr$^{3+}$:Y$_2$SiO$_5$ showing $T_2^\textrm{start}$ without ZEFOZ and $T_2^\textrm{opt}$ with ZEFOZ. Figures (b), (d) and (e) were reproduced from refs \cite{Shoudu1999, Longdell2006, Heinze2014}, respectively, with permission from the American Physical Society and Elsevier. Data points for figure (c) were extracted from refs \cite{Yano1991, Zhong2015, Bottger2016b, equallHomogeneousBroadeningHyperfine1995, Bottger2006}. }
        \label{fig:Lanthanide_Figure}
    \end{figure*}

Lanthanide spin centres benefit from highly shielded f-orbital electrons compared to d-orbital systems, leading to relatively long coherence times at low temperatures \cite{Sorace2015}. The core-like nature of f-orbitals also means, compared to d-orbital systems, that the ligand-field environment has a much weaker influence on state energies (a few THz), compared to electron-electron interactions ($\approx$ 800 - 1600 nm) and spin-orbit coupling ($\approx$ 10s THz) \cite{Jiang2014, Parker2020} (see Figure \ref{fig:Lanthanide_Figure}a). Lanthanide electron spins can also exhibit strong hyperfine coupling (tens of MHz) with I=5/2 (Eu, Pr, $^{173}$Yb), I=1/2 ($^{171}$Yb), I=7/2 ($^{149}$Sm, $^{167}$Er) nuclei, leading to a diverse spin-optical addressibility schemes that can be further enhanced using applied magnetic fields. Moreover, strong spin-orbit coupling can lead to high magnetic anisotropy that can protect spin states from small magnetic fluctuations \cite{Gaita-Arino2019, Aromi2019}. Examples of optically active materials benefit from narrow and stable spin-dependent emission profiles at near-infrared frequencies, making them potentially compatible with conventional telecom fibre optics. Colour centres for which optical addressability has been established include Ce$^{3+}$ \cite{Liang2017}, Eu$^{3+}$- \cite{Yano1991}, Pr$^{3+}$-\cite{equallHomogeneousBroadeningHyperfine1995}, Er$^{3+}$- \cite{Rancic2018}, Yb$^{3+}$- \cite{Ortu2018}, and Sm$^{3+}$:Y$_2$SiO$_5$ \cite{Jobbitt2021} (Figure \ref{fig:Lanthanide_Figure}c). 

In the solid-state, Y$_2$SiO$_5$ has been favoured as host matrix due to the ability to grow large crystals with excellent optical properties using the Czochralski method \cite{Shoudu1999} (Figure \ref{fig:Lanthanide_Figure}b). However, as the only naturally occurring isotope, $^{89}$Y harbours an I=1/2 nuclear spin that ultimately limits decoherence times. To reduce the impact of the spin bath, various decoupling techniques have emerged \cite{Fraval2004, Fraval2005, Zhong2019}. Perhaps the most successful is the so-called ``ZEFOZ'' technique, whereby a magnetic field is applied such that the magnetic-field dependence of the spin-transition frequency is very close to zero \cite{McAuslan2012} (Figure \ref{fig:Lanthanide_Figure}d). In this ``clock transition'' regime, spins are first-order insensitive to small fluctuations in local magnetic fields. Using this technique with Eu$^{3+}$:Y$_2$SiO$_5$, Zhong \textit{et al.,} demonstrated it is possible to acquire $T_2$ $\gg$100~ms at 2~K, where spin-phonon coupling is negligible. Remarkably, combined with dynamic decoupling methods, $T_\textrm{DD}$ was measured to be 370$\pm$60~mins, reaching a critical milestone whereby the distance-dependent decoherence becomes less for spin-transport than it is during light-transport of quantum states \cite{Zhong2015}.

Interestingly, the larger magnetic moment of Pr$^{3+}$ can give rise to ``frozen core''-type behaviour, whereby local Y-spins become dephased from the bulk crystal and hence exhibit reduced dephasing influence on the Pr$^{3+}$ spins \cite{Fraval2004}. Equall \textit{et al.,} measured homogenous field-dependent and crystal structure site-dependent linewidths between 2.5 and 0.85~kHz, and a corresponding $T_2$ as high as 377~$\mu$s at 1.4~K \cite{equallHomogeneousBroadeningHyperfine1995}. Combined with the ZEFOZ method, the $T_2$ can reach 82~ms at 1.5~K \cite{Fraval2004} (see example Figure \ref{fig:Lanthanide_Figure}e) and with further dynamical decoupling $T_\textrm{DD}$ up to 1~min can be achieved, approaching the population lifetime limit \cite{Heinze2014}. Coherence times can also be enhanced using a host matrix whereby the principal host ion (e.g., Y) has a more closely matched ionic radius to the dopant, leading to reduced crystallographic distortions. For example, in a Pr$^{3+}$:La$_2$(WO$_4$)$_3$ system a $T_2$ of 158$\pm$7 ms has been measured at $\approx$ 4~K\cite{Lovric2011}.

Spin coherence times of Er$^{3+}$:Y$_2$SiO$_5$ have so far been shorter than the best performing Eu$^{3+}$ materials with $T_2$ measured at 1.3$\pm$0.01 seconds at 1.4 K, despite exhibiting the frozen core effect \cite{Rancic2018}. However, this was achieved without using the ZEFOZ method, and therefore, these times can likely be significantly extended. Er$^{3+}$ has also been studied in a Y$_2$O$_3$ host with $T_2$ reaching $\approx$ 140$~\mu$s at 1.8~K and with an applied field of 4~T
\cite{bottgerOpticalDecoherenceSpectroscopy2024}. Decoherence was dominated by phonon-driven dipole-dipole interactions and the nuclear spin bath at high fields, similar to Er$^{3+}$:KTiOPO$_4$ where $T_2$ was measured at $\approx \,$200~$\mu$s under similar conditions \cite{Bottger2016}.

$^{171}$Yb$^{3+}$ holds a unique position amongst the lanthanide ions discussed so far due to its S=1/2 and I=1/2 electron spin and hyperfine structure, resulting in a simple 4-level system. Of the $^{171}$Yb$^{3+}$-doped materials \cite{Bottger2016b}, $^{171}$Yb$^{3+}$:Y$_2$SiO$_5$ appears to exhibit the longest spin coherence times. This material was initially studied by X-band EPR spectroscopy and presented with an electron $T_1$ of $\approx$ 5 seconds at 2.5 K with an applied field of $\approx$ 90 mT. $T_1$ quickly increases above $\approx$ 4 K due to Raman relaxation where $T_2$ becomes ultimately limited by $T_1$\cite{Lim2018}. At 2.5~K, $T_2$ was optimised at $\approx$ 1~T to 73~$\mu$s and improved further by dynamical decoupling reaching 550~$\mu$s. In the same experiments, the authors record nuclear $T_1$ and $T_2$ at 4.5~K of 4 and 0.35~ms, respectively. Using an optical approach, Ortu \textit{et al.,} improved the coherent properties of $^{171}$Yb$^{3+}$:Y$_2$SiO$_5$ by employing the ZEFOZ method such that the electron $T_2$ remains above 100 $\mu$s at 5.6~K and the nuclear $T_2$ extends to 1~ms \cite{Ortu2018}. Using a different approach, Welinski \textit{et al.,} demonstrated that coherence can also be extended by first polarising host nuclear spins through spin diffusion. At 2~K, the authors first excite $^{171}$Yb spins before allowing them to equilibrate over a few seconds through spectral diffusion over the inhomogeneous linewidth. The result is an effective ``hole burning'' in the absorption spectrum of $^{171}$Yb$^{3+}$:Y$_2$SiO$_5$ and up to 90$\%$ nuclear spin polarisation, thereby effectively generating mK spin temperatures and improving the optical $T_2$ from 0.3 to 0.8~ms\cite{Welinski2020}. 

To our knowledge, pulsed optical decoherence studies have not been performed on Sm$^{3+}$:Y$_2$SiO$_5$, however, its I=7/2 nucleus and strong hyperfine coupling may be useful for qudit systems, and it is also predicted to be less sensitive to magnetic field fluctuations than Er$^{3+}$ \cite{Jobbitt2021, Jobbitt2022}.

Finally, in recent years, Ce$^{3+}$:YAG has become a well studied example of a potential lanthanide spin material that is readily available. As with the previous V$^{4+}$ and Mo$^{2+}$ spin centres found in SiC, Ce$^{3+}$ exhibits a S=1/2 ground state and is not initialised through ISC, but rather uniquely uses left- or right-handed circularly polarised light to selectively depopulate a spin-dependent ZPL \cite{Kolesov2013}. Moreover, unlike many lanthanide defects, this 4f- to 5d-orbital transition exhibits a high oscillator strength and can be excited using green light with yellow emission, generating a theoretical maximum of 99.7$\%$ spin polarisation in its excited state. This state only lasts $\approx$ 63 ns or so before relaxing back to the ground state with the a reported ODMR contrast of 98$\%$ \cite{Xia2015}. Whilst maintenance of robust spin properties still requires low temperatures, in part due to the abundance of $^{27}$Al (I=5/2) in the YAG \cite{Belykh2022, Azamat2017}, detectable spin coherence is maintained up to 20 K - significantly higher than other lanthanides, and $T_1$ can even be measured at room temperature \cite{Kolesov2013}. Cross relaxation of commonly found co-impurities has been used to optically-read out the spin-state of species like Tb$^{3+}$ and Gb$^{3+}$ \cite{Tolmachev2017, Edinach2019}, demonstrating the capacity of Ce$^{3+}$ to act as a spin read-out relay for otherwise dark spins - an essential property for nanoscale sensing. Moreover, Ce$^{3+}$:YAG may even demonstrate improved ODMR spin properties when manufactured into thin-films, making it a promising platform for incorporation into devices \cite{Chai2025}.




\section{Optimising Performance of Spin-Qubit Materials }

\begin{figure}[ht]
        \centering
        \includegraphics[width=\columnwidth]{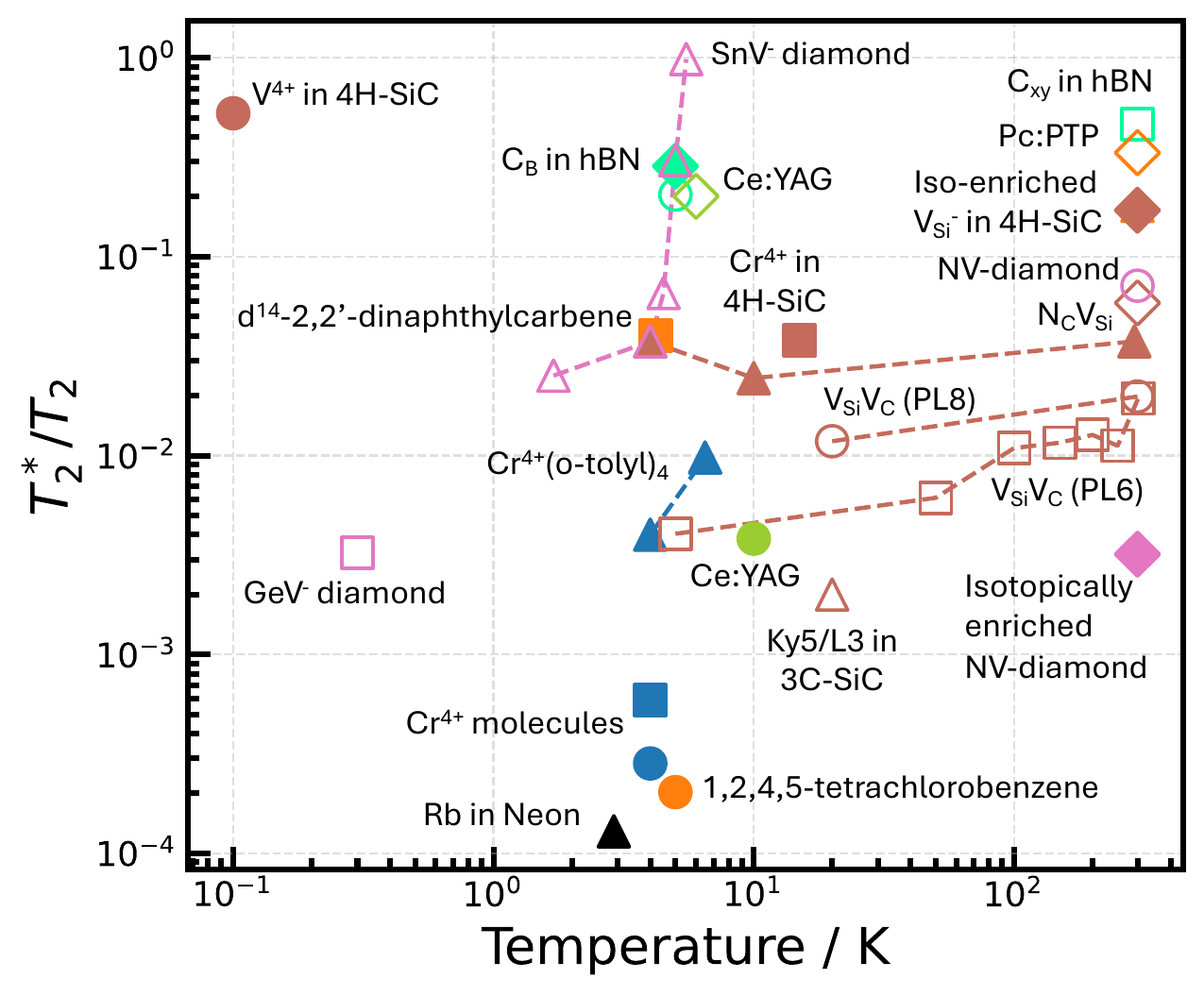}
        \caption{\textbf{Comparison of $T_2$ and $T_2^*$ for different spin systems as a measure of inhomogeneity.}}
        \label{fig:t2vt2star}
    \end{figure}

From our review, it is clear that spin-based qubit candidates demonstrate potential in several fields of quantum technology. However, unsurprisingly, their spin coherence lifetimes are limited by temperature and the spin-bath dependence of the spin properties and optical contrast between different spin states. Across all material platforms, there is significant magnetic inhomogeneity that emerges from random local spin environments. To understand the extent to which inhomogeneity infects different materials, it could be instructive to consider the ratio of $T_2^*$/$T_2$ (Figure \ref{fig:t2vt2star}, the so-called ``inhomogeneity parameter''). Hence, a low ratio indicates that $T_2$ is close to $T_2^*$ and the materials are limited by dynamic, not static, sources of decoherence. Using the available data where $T_2^*$ and $T_2$ were measured under similar conditions, there appears to emerge a distinct advantage for molecules and van der Waals materials at higher temperatures despite their lack of isotopic enrichment. This likely stems from the use of molecular crystals and the inherent 2D-order associated with materials such as hBN, which helps to ensure that all molecules ``feel'' the same magnetic environment. Interestingly, despite the significant difficulties associated with synthesising aligned 3D-spin defects in diamond, SnV$^-$diamond (measured as single spins) exhibits the lowest ratio at low temperatures. Considering its robust quantum spin parameters, SnV$^-$diamond could be a leading candidate for low-temperature applications. 

Spin parameters can be improved by positional engineering of defect centres. Clustering (or the straggling) of spin-active dopants in a substrate poses spin-spin coupling from the environment (nuclear spins), which decreases $T_2$. Controlled doping becomes crucial in decreasing spin density around spin-active defect centres. Eliminating unwanted spins like nuclear spins requires isotopic purity of substrate material or the careful doping of spin-active centres or defects within the host matrix. More recently, Plasma-Enhanced CVD methods have been used to achieve higher deposition rates while minimising the impact of energetic ions or electrons affecting the colour centres in NV and SnV diamonds \cite{ZhangPECVD2016}. Another novel method to precisely control the position of each defect is laser writing. Aberration-corrected optics allow for the precise positioning of vacancies in diamond systems, with a 45$\%$ success probability of a vacancy being located within 200 nm of a desired position \cite{Chen2016}.

Due to their ability to effectively engineer the placement of molecules in 3D chemical systems, materials with charge transfer or hydrogen bonding motifs may yet demonstrate advantages. Moreover, as seen with Pc:PTP, and Pr$^{3+}$:Y$_2$SiO$_5$ vs.~the La$_2$(WO$_5$)$_3$ host, material engineers should avoid defect-site strain by selecting hosts with closely matched physical parameters to the defect. Finally, further improvements can be realised through dynamical decoupling methods that are tailored for each application. For example, while engineering clock transitions using magnetic fields or ZFS is appealing for quantum optics where spectral stability is prized, it is not necessarily useful for sensing or information processing. This is because the associated reduced magnetic field sensitivity would reduce the quantum operation fidelity. On the other hand, focused electromagnetic driving of parasitic impurities, such as N-centres in diamond, has yet to be significantly explored in molecular systems. Significant improvements in $T_2$ and $T_2^*$ would likely emerge from a combination of field-driving and pulsed refocusing methods such as CPMG.




\section{Conclusion and Outlook}

Optically addressable spin systems show great potential as a diverse form of qubit media for quantum sensing, communications, and information processing applications. From this review, we have attempted to benchmark the different families of materials and identify useful investigative and experimental approaches that can be translated across the field. The most significant hurdles faced by chemists and materials engineers remain the strong temperature dependence of the spin-lattice relaxation and thermal polarisation of the spin bath, which renders most heavy atom-containing systems impractical above a few kelvin, but below which these materials demonstrate the longest coherence times by a significant margin. However, above liquid helium temperatures, it is clear that light-element materials such as colour centres in diamond, SiC and most recently, molecular systems demonstrate more robust quantum spin parameters. Nevertheless, significant advancements in dynamic decoupling techniques within the last 20 years and isotope engineering have enabled the realisation of remarkably competitive quantum spin properties. Within the next decade, it should be expected that further advances and knowledge transfer between investigators of different material platforms will lead to devices capable of significantly impacting society, especially in the field of magnetic field sensors and coherent quantum optics. In particular, materials that are currently prized for their spectral stability and employed mainly for applications in quantum optics and communication or networking (e.g., f-block, heavy-atom vacancy centres) will benefit from incorporating the relatively weak phononic sensitivity and high brightness found in materials primarily used for quantum sensing and information processing (e.g., defects in diamond, SiC, hBN and molecules), and vice versa. We also hope that by collating (meta)data on the available spin qubit candidates that the field will eventually benefit from the processing and predictive power of machine learning and artificial intelligence approaches to materials discovery.  

\section{Acknowledgments}

This work was supported by the UK Engineering and Physical Science Research Council through grants EP/V048430/1 and EP/W027542/1, and the Department of Materials, Imperial College London. The authors are also very grateful to Dr. Daan M. Arroo for fruitful discussions and advice.

\section{Author contributions}

CAT and MA curated the data. CAT, MO and MA contributed to the writing of the manuscript.

\bibliographystyle{unsrt}  
\bibliography{references}  

@article{arecchiAtomicCoherentStates1972,
  title = {Atomic {{Coherent States}} in {{Quantum Optics}}},
  author = {Arecchi, F. T. and Courtens, Eric and Gilmore, Robert and Thomas, Harry},
  date = {1972-12-01},
  journal = {Physical Review A},
  shortjournal = {Phys. Rev. A},
  volume = {6},
  number = {6},
  pages = {2211--2237},
  issn = {0556-2791},
  doi = {10.1103/PhysRevA.6.2211},
  url = {https://link.aps.org/doi/10.1103/PhysRevA.6.2211},
  urldate = {2023-12-01},
  langid = {english}, year = {1972}
}

@article{castelvecchiIBMReleasesFirstever2023,
  title = {{{IBM}} Releases First-Ever 1,000-Qubit Quantum Chip},
  author = {Castelvecchi, Davide},
  date = {2023-12-04},
  journal = {Nature},
  publisher = {{Nature Publishing Group}},
  doi = {10.1038/d41586-023-03854-1},
  url = {https://www.nature.com/articles/d41586-023-03854-1},
  urldate = {2023-12-04},
  langid = {english},
  year = {2023}
}

@misc{blackSupercomputersQuantumMachines2023,
  title = {This German startup wants to build portable quantum computers using diamonds - and says its QPU will sit next to a GPU or a CPU one day. Article available: https://quantumbrilliance.com/news-media},
  author = {Quantum Brilliance},
  date = {2025-06-23},
  url = {https://quantumbrilliance.com/news-media},
  urldate = {2025-06-23},
  langid = {american},
  year = {2025}
}

@article{baylissEnhancingSpinCoherence2022,
  title = {Enhancing {{Spin Coherence}} in {{Optically Addressable Molecular Qubits}} through {{Host-Matrix Control}}},
  author = {Bayliss, S. L. and Deb, P. and Laorenza, D. W. and Onizhuk, M. and Galli, G. and Freedman, D. E. and Awschalom, D. D.},
  date = {2022-08-18},
  journal = {Physical Review X},
  shortjournal = {Phys. Rev. X},
  volume = {12},
  number = {3},
  pages = {031028},
  issn = {2160-3308},
  doi = {10.1103/PhysRevX.12.031028},
  url = {https://link.aps.org/doi/10.1103/PhysRevX.12.031028},
  urldate = {2024-06-28},
  langid = {english}, year = {2022}
}

@article{Bottger2006,
author = {B{\"{o}}ttger, Thomas and Sun, Y. and Thiel, C. W. and Cone, R. L.},
doi = {10.1103/PhysRevB.74.075107},
issn = {1098-0121},
journal = {Physical Review B},
month = {Aug},
number = {7},
pages = {075107},
title = {{Spectroscopy and dynamics of Er$^{3+}$:Y$_2$SiO$_5$ at 1.5 $\mu$m}},
url = {https://link.aps.org/doi/10.1103/PhysRevB.74.075107},
volume = {74},
year = {2006}
}

@article{bottgerOpticalDecoherenceSpectroscopy2024,
  title = {Optical Decoherence, Spectroscopy, and Magnetic g Tensors for the 1.5-um $^{4}${{I}}$_{15/2}$ - $^{4}${{I}}$_{13/2}$ transitions of {{Er}}$^{3+}$ dopants at the {{C}}$_2$-symmetry Site in {{Y}}$_2${{O}}$_3$},
  author = {Böttger, Thomas and Harris, T. L. and Reinemer, G. D. and Thiel, C. W. and Cone, R. L.},
  date = {2024-07-17},
  journal = {Physical Review B},
  shortjournal = {Phys. Rev. B},
  volume = {110},
  number = {4},
  pages = {045132},
  issn = {2469-9950, 2469-9969},
  doi = {10.1103/PhysRevB.110.045132},
  url = {https://link.aps.org/doi/10.1103/PhysRevB.110.045132},
  urldate = {2024-07-20},
  langid = {english}, year = {2024}
}

@article{dargyteOpticalSpincoherenceProperties2021,
  title = {Optical and Spin-Coherence Properties of Rubidium Atoms Trapped in Solid Neon},
  author = {Dargyte, Ugne and Lancaster, David M. and Weinstein, Jonathan D.},
  date = {2021-09-21},
  journal = {Physical Review A},
  shortjournal = {Phys. Rev. A},
  volume = {104},
  number = {3},
  pages = {032611},
  issn = {2469-9926, 2469-9934},
  doi = {10.1103/PhysRevA.104.032611},
  url = {https://link.aps.org/doi/10.1103/PhysRevA.104.032611},
  urldate = {2024-07-09},
  langid = {english}, year = {2021}
}

@article{DiVincenzo2000,
   author = {David P. DiVincenzo},
   doi = {10.1002/1521-3978(200009)48:9/11<771::AID-PROP771>3.0.CO;2-E},
   issn = {00158208},
   issue = {9-11},
   journal = {Fortschritte der Physik},
   month = {9},
   pages = {771-783},
   publisher = {Wiley-VCH Verlag},
   title = {The Physical Implementation of Quantum Computation},
   volume = {48},
   url = {https://onlinelibrary.wiley.com/doi/10.1002/1521-3978(200009)48:9/11<771::AID-PROP771>3.0.CO;2-E},
   year = {2000}
}

@article{divincenzoQuantumComputersQuantum1999,
  title = {Quantum Computers and Quantum Coherence},
  author = {DiVincenzo, David P and Loss, Daniel},
  date = {1999-10},
  journal = {Journal of Magnetism and Magnetic Materials},
  shortjournal = {Journal of Magnetism and Magnetic Materials},
  volume = {200},
  number = {1-3},
  pages = {202--218},
  issn = {03048853},
  doi = {10.1016/S0304-8853(99)00315-7},
  url = {https://linkinghub.elsevier.com/retrieve/pii/S0304885399003157},
  urldate = {2023-11-07},
  langid = {english}, year = {1999}
}

@article{equallHomogeneousBroadeningHyperfine1995,
  title = {Homogeneous Broadening and Hyperfine Structure of Optical Transitions in {{Pr}}$^{3+}$:{{Y}}$_2${{Si}}{{O}}$_5$},
  shorttitle = {Homogeneous Broadening and Hyperfine Structure of Optical Transitions in {{Pr}} 3 +},
  author = {Equall, R. W. and Cone, R. L. and Macfarlane, R. M.},
  date = {1995-08-01},
  journal = {Physical Review B},
  shortjournal = {Phys. Rev. B},
  volume = {52},
  number = {6},
  pages = {3963--3969},
  issn = {0163-1829, 1095-3795},
  doi = {10.1103/PhysRevB.52.3963},
  url = {https://link.aps.org/doi/10.1103/PhysRevB.52.3963},
  urldate = {2024-07-20},
  langid = {english}, year = {1995}
}

@article{sternQuantumCoherentSpin,
author = {Stern, Hannah L and {M. Gilardoni}, Carmem and Gu, Qiushi and {Eizagirre Barker}, Simone and Powell, Oliver F J and Deng, Xiaoxi and Fraser, Stephanie A and Follet, Louis and Li, Chi and Ramsay, Andrew J and Tan, Hark Hoe and Aharonovich, Igor and Atat{\"{u}}re, Mete},
doi = {10.1038/s41563-024-01887-z},
issn = {1476-1122},
journal = {Nature Materials},
month = {Oct},
number = {10},
pages = {1379--1385},
title = {{A quantum coherent spin in hexagonal boron nitride at ambient conditions}},
url = {https://www.nature.com/articles/s41563-024-01887-z},
volume = {23},
year = {2024}
}

@article{Gong_2023,
author = {Gong, Ruotian and He, Guanghui and Gao, Xingyu and Ju, Peng and Liu, Zhongyuan and Ye, Bingtian and Henriksen, Erik A. and Li, Tongcang and Zu, Chong},
doi = {10.1038/s41467-023-39115-y},
eprint = {2210.11485},
issn = {2041-1723},
journal = {Nature Communications},
month = {Jun},
number = {1},
pages = {3299},
pmid = {37280252},
publisher = {Springer US},
title = {{Coherent dynamics of strongly interacting electronic spin defects in hexagonal boron nitride}},
url = {https://www.nature.com/articles/s41467-023-39115-y},
volume = {14},
year = {2023}
}

@article{gorgonReversibleSpinopticalInterface2023a,
  title = {Reversible Spin-Optical Interface in Luminescent Organic Radicals},
  author = {Gorgon, Sebastian and Lv, Kuo and Grüne, Jeannine and Drummond, Bluebell H. and Myers, William K. and Londi, Giacomo and Ricci, Gaetano and Valverde, Danillo and Tonnelé, Claire and Murto, Petri and Romanov, Alexander S. and Casanova, David and Dyakonov, Vladimir and Sperlich, Andreas and Beljonne, David and Olivier, Yoann and Li, Feng and Friend, Richard H. and Evans, Emrys W.},
  date = {2023-08},
  journal = {Nature},
  volume = {620},
  number = {7974},
  pages = {538--544},
  publisher = {Nature Publishing Group},
  issn = {1476-4687},
  doi = {10.1038/s41586-023-06222-1},
  url = {https://www.nature.com/articles/s41586-023-06222-1},
  urldate = {2024-08-18},
  langid = {english}, year = {2023}
}

@article{gottschollRoomTemperatureCoherent2021,
  title = {Room Temperature Coherent Control of Spin Defects in Hexagonal Boron Nitride},
  author = {Gottscholl, Andreas and Diez, Matthias and Soltamov, Victor and Kasper, Christian and Sperlich, Andreas and Kianinia, Mehran and Bradac, Carlo and Aharonovich, Igor and Dyakonov, Vladimir},
  date = {2021-04-02},
  journal = {Science Advances},
  volume = {7},
  number = {14},
  pages = {eabf3630},
  publisher = {{American Association for the Advancement of Science}},
  doi = {10.1126/sciadv.abf3630},
  url = {https://www.science.org/doi/10.1126/sciadv.abf3630},
  urldate = {2023-12-07}, year = {2021}
}

@article{healeyOptimisationElectronIrradiation2024,
author = {Healey, Alexander J. and Singh, Priya and Robertson, Islay O. and Gavin, Christopher and Scholten, Sam C. and Broadway, David A. and Reineck, Philipp and Abe, Hiroshi and Ohshima, Takeshi and Kianinia, Mehran and Aharonovich, Igor and Tetienne, Jean-Philippe},
doi = {10.1088/2633-4356/ad65ae},
issn = {2633-4356},
journal = {Materials for Quantum Technology},
month = {Sep},
number = {3},
pages = {035701},
title = {{Optimisation of electron irradiation for creating spin ensembles in hexagonal boron nitride}},
url = {https://iopscience.iop.org/article/10.1088/2633-4356/ad65ae},
volume = {4},
year = {2024}
}

@article{koehlResonantOpticalSpectroscopy2017,
  title = {Resonant Optical Spectroscopy and Coherent Control of {{C}}r$^{4+}$ Spin Ensembles in {{SiC}} and {{GaN}}},
  author = {Koehl, William F. and Diler, Berk and Whiteley, Samuel J. and Bourassa, Alexandre and Son, N. T. and Janzén, Erik and Awschalom, David D.},
  date = {2017-01-19},
  journal = {Physical Review B},
  shortjournal = {Phys. Rev. B},
  volume = {95},
  number = {3},
  pages = {035207},
  issn = {2469-9950, 2469-9969},
  doi = {10.1103/PhysRevB.95.035207},
  url = {https://link.aps.org/doi/10.1103/PhysRevB.95.035207},
  urldate = {2024-06-26},
  langid = {english}, year = {2017}
}

@article{Chizzini2022,
author = {Chizzini, Mario and Crippa, Luca and Zaccardi, Luca and Macaluso, Emilio and Carretta, Stefano and Chiesa, Alessandro and Santini, Paolo},
doi = {10.1039/d2cp01228f},
issn = {14639076},
journal = {Physical Chemistry Chemical Physics},
number = {34},
pages = {20030--20039},
pmid = {35833847},
publisher = {Royal Society of Chemistry},
title = {{Quantum error correction with molecular spin qudits}},
volume = {24},
year = {2022}
}

@article{linTemperatureDependenceDivacancy2021,
  title = {Temperature Dependence of Divacancy Spin Coherence in Implanted Silicon Carbide},
  author = {Lin, Wu-Xi and Yan, Fei-Fei and Li, Qiang and Wang, Jun-feng and Hao, Zhi-He and Zhou, Ji-Yang and Li, Hao and You, Li-Xing and Xu, Jin-Shi and Li, Chuan-Feng and Guo, Guang-Can},
  date = {2021-09-15},
  journal = {Physical Review B},
  shortjournal = {Phys. Rev. B},
  volume = {104},
  number = {12},
  pages = {125305},
  issn = {2469-9950, 2469-9969},
  doi = {10.1103/PhysRevB.104.125305},
  url = {https://link.aps.org/doi/10.1103/PhysRevB.104.125305},
  urldate = {2023-11-29},
  langid = {english}, year = {2021}
}

@article{liuCoherentQuantumControl2019,
  title = {Coherent Quantum Control of Nitrogen-Vacancy Center Spins near 1000 Kelvin},
  author = {Liu, Gang-Qin and Feng, Xi and Wang, Ning and Li, Quan and Liu, Ren-Bao},
  date = {2019-03-22},
  journal = {Nature Communications},
  shortjournal = {Nat Commun},
  volume = {10},
  number = {1},
  pages = {1344},
  publisher = {Nature Publishing Group},
  issn = {2041-1723},
  doi = {10.1038/s41467-019-09327-2},
  url = {https://www.nature.com/articles/s41467-019-09327-2},
  urldate = {2024-08-12},
  langid = {english}, year = {2019}
}

@article{ngMoveAsidePentacene2022,
author = {Ng, Wern and Xu, Xiaotian and Attwood, Max and Wu, Hao and Meng, Zhu and Chen, Xi and Oxborrow, Mark},
doi = {10.1002/adma.202300441},
eprint = {2211.06176},
issn = {0935-9648},
journal = {Advanced Materials},
month = {Jun},
number = {22},
pages = {2300441},
pmid = {36919948},
title = {{Move Aside Pentacene: Diazapentacene‐Doped para‐Terphenyl, a Zero‐Field Room‐Temperature Maser with Strong Coupling for Cavity Quantum Electrodynamics}},
url ={https://onlinelibrary.wiley.com/doi/10.1002/adma.202300441},
volume = {35},
year = {2023}
}

@article{pinoDemonstrationTrappedionQuantum2021,
  title = {Demonstration of the Trapped-Ion Quantum {{CCD}} Computer Architecture},
  author = {Pino, J. M. and Dreiling, J. M. and Figgatt, C. and Gaebler, J. P. and Moses, S. A. and Allman, M. S. and Baldwin, C. H. and Foss-Feig, M. and Hayes, D. and Mayer, K. and Ryan-Anderson, C. and Neyenhuis, B.},
  date = {2021-04},
  journal = {Nature},
  shortjournal = {Nature},
  volume = {592},
  number = {7853},
  pages = {209--213},
  publisher = {{Nature Publishing Group}},
  issn = {1476-4687},
  doi = {10.1038/s41586-021-03318-4},
  url = {https://www.nature.com/articles/s41586-021-03318-4},
  urldate = {2023-12-03},
  issue = {7853},
  langid = {english},
  year = {2021}
}

@article{rosenthalMicrowaveSpinControl2023,
  title = {Microwave {{Spin Control}} of a {{Tin-Vacancy Qubit}} in {{Diamond}}},
  author = {Rosenthal, Eric I. and Anderson, Christopher P. and Kleidermacher, Hannah C. and Stein, Abigail J. and Lee, Hope and Grzesik, Jakob and Scuri, Giovanni and Rugar, Alison E. and Riedel, Daniel and Aghaeimeibodi, Shahriar and Ahn, Geun Ho and Van Gasse, Kasper and Vučković, Jelena},
  date = {2023-08-30},
  journal = {Physical Review X},
  shortjournal = {Phys. Rev. X},
  volume = {13},
  number = {3},
  pages = {031022},
  issn = {2160-3308},
  doi = {10.1103/PhysRevX.13.031022},
  url = {https://link.aps.org/doi/10.1103/PhysRevX.13.031022},
  urldate = {2024-08-02},
  langid = {english}, year = {2023}
}

@article{Scholten2024,
archivePrefix = {arXiv},
arxivId = {2306.16600},
author = {Scholten, Sam C. and Singh, Priya and Healey, Alexander J. and Robertson, Islay O. and Haim, Galya and Tan, Cheng and Broadway, David A. and Wang, Lan and Abe, Hiroshi and Ohshima, Takeshi and Kianinia, Mehran and Reineck, Philipp and Aharonovich, Igor and Tetienne, Jean-Philippe},
doi = {10.1038/s41467-024-51129-8},
eprint = {2306.16600},
issn = {2041-1723},
journal = {Nature Communications},
month = {Aug},
number = {1},
pages = {6727},
pmid = {39112477},
publisher = {Springer US},
title = {{Multi-species optically addressable spin defects in a van der Waals material}},
url = {http://dx.doi.org/10.1038/s41467-024-51129-8 https://www.nature.com/articles/s41467-024-51129-8},
volume = {15},
year = {2024}
}

@article{Singh2025,
archivePrefix = {arXiv},
arxivId = {2402.13898},
author = {Singh, Harpreet and D'Souza, Noella and Zhong, Keyuan and Druga, Emanuel and Oshiro, Julianne and Blankenship, Brian and Montis, Riccardo and Reimer, Jeffrey A. and Breeze, Jonathan D. and Ajoy, Ashok},
doi = {10.1103/PhysRevResearch.7.013192},
eprint = {2402.13898},
issn = {2643-1564},
journal = {Physical Review Research},
month = {Feb},
number = {1},
pages = {013192},
publisher = {American Physical Society},
title = {{Room-temperature quantum sensing with photoexcited triplet electrons in organic crystals}},
url = {http://arxiv.org/abs/2402.13898 https://link.aps.org/doi/10.1103/PhysRevResearch.7.013192},
volume = {7},
year = {2025}
}

@article{Mena2024,
author = {Mena, Adrian and Mann, Sarah K. and Cowley-Semple, Angus and Bryan, Emma and Heutz, Sandrine and McCamey, Dane R. and Attwood, Max and Bayliss, Sam L.},
doi = {10.1103/PhysRevLett.133.120801},
issn = {0031-9007},
journal = {Physical Review Letters},
month = {Sep},
number = {12},
pages = {120801},
title = {{Room-Temperature Optically Detected Coherent Control of Molecular Spins}},
url = {https://link.aps.org/doi/10.1103/PhysRevLett.133.120801},
volume = {133},
year = {2024}
}

@article{takahashiQuenchingSpinDecoherence2008,
  title = {Quenching {{Spin Decoherence}} in {{Diamond}} through {{Spin Bath Polarization}}},
 author = {Takahashi, Susumu and Hanson, Ronald and van~Tol, Johan and Sherwin, Mark S. and Awschalom, David D.},
  date = {2008-07-23},
  journal = {Physical Review Letters},
  shortjournal = {Phys. Rev. Lett.},
  volume = {101},
  number = {4},
  pages = {047601},
  publisher = {{American Physical Society}},
  doi = {10.1103/PhysRevLett.101.047601},
  url = {https://link.aps.org/doi/10.1103/PhysRevLett.101.047601},
  urldate = {2023-11-27}, year = {2008}
}

@article{Weber2010,
  title = {Quantum computing with defects},
  volume = {107},
  ISSN = {1091-6490},
  url = {http://dx.doi.org/10.1073/pnas.1003052107},
  DOI = {10.1073/pnas.1003052107},
  number = {19},
  journal = {Proceedings of the National Academy of Sciences},
  publisher = {Proceedings of the National Academy of Sciences},
  author = {Weber,  J. R. and Koehl,  W. F. and Varley,  J. B. and Janotti,  A. and Buckley,  B. B. and Van de Walle,  C. G. and Awschalom,  D. D.},
  year = {2010},
  month = apr,
  pages = {8513–8518}
}

@article{wolfowiczVanadiumSpinQubits2020,
  title = {Vanadium Spin Qubits as Telecom Quantum Emitters in Silicon Carbide},
  author = {Wolfowicz, Gary and Anderson, Christopher P. and Diler, Berk and Poluektov, Oleg G. and Heremans, F. Joseph and Awschalom, David D.},
  date = {2020-05},
  journal = {Science Advances},
  volume = {6},
  number = {18},
  pages = {eaaz1192},
  publisher = {American Association for the Advancement of Science},
  doi = {10.1126/sciadv.aaz1192},
  url = {https://www.science.org/doi/10.1126/sciadv.aaz1192},
  urldate = {2024-07-12}, year = {2020}
}

@article{ZhangPECVD2016,
  title = {Templated growth of diamond optical resonators via plasma-enhanced chemical vapor deposition},
  volume = {109},
  ISSN = {1077-3118},
  url = {http://dx.doi.org/10.1063/1.4961536},
  DOI = {10.1063/1.4961536},
  number = {8},
  journal = {Applied Physics Letters},
  publisher = {AIP Publishing},
  author = {Zhang,  X. and Hu,  E. L.},
  year = {2016},
  month = aug 
}

@article{Mrozek2015,
author = {Mr{\'{o}}zek, Mariusz and Rudnicki, Daniel and Kehayias, Pauli and Jarmola, Andrey and Budker, Dmitry and Gawlik, Wojciech},
doi = {10.1140/epjqt/s40507-015-0035-z},
issn = {2196-0763},
journal = {EPJ Quantum Technology},
month = {Dec},
number = {1},
pages = {22},
title = {{Longitudinal spin relaxation in nitrogen-vacancy ensembles in diamond}},
url = {http://epjquantumtechnology.springeropen.com/articles/10.1140/epjqt/s40507-015-0035-z},
volume = {2},
year = {2015}
}

@article{Cambria2023,
author = {Cambria, M. C. and Norambuena, A. and Dinani, H. T. and Thiering, G. and Gardill, A. and Kemeny, I. and Li, Y. and Lordi, V. and Gali, {\'{A}}. and Maze, J. R. and Kolkowitz, S.},
doi = {10.1103/PhysRevLett.130.256903},
issn = {0031-9007},
journal = {Physical Review Letters},
publisher = {Physical Review Letters},
month = {Jun},
number = {25},
pages = {256903},
title = {{Temperature-Dependent Spin-Lattice Relaxation of the Nitrogen-Vacancy Spin Triplet in Diamond}},
url = {https://link.aps.org/doi/10.1103/PhysRevLett.130.256903},
volume = {130},
year = {2023}
}

@article{Atzori2019,
   author = {Matteo Atzori and Roberta Sessoli},
   doi = {10.1021/jacs.9b00984},
   issn = {15205126},
   issue = {29},
   journal = {Journal of the American Chemical Society},
   pages = {11339-11352},
   pmid = {31287678},
   title = {The Second Quantum Revolution: Role and Challenges of Molecular Chemistry},
   volume = {141},
   year = {2019},
}

@article{Wasielewski2020,
   author = {Michael R. Wasielewski and Malcolm D. E. Forbes and Natia L. Frank and Karol Kowalski and Gregory D. Scholes and Joel Yuen-Zhou and Marc A. Baldo and Danna E. Freedman and Randall H. Goldsmith and Theodore Goodson and Martin L. Kirk and James K. McCusker and Jennifer P. Ogilvie and David A. Shultz and Stefan Stoll and K. Birgitta Whaley},
   doi = {10.1038/s41570-020-0200-5},
   issn = {2397-3358},
   issue = {9},
   journal = {Nature Reviews Chemistry},
   month = {9},
   pages = {490-504},
   publisher = {Springer US},
   title = {Exploiting chemistry and molecular systems for quantum information science},
   volume = {4},
   url = {http://dx.doi.org/10.1038/s41570-020-0200-5 http://www.nature.com/articles/s41570-020-0200-5},
   year = {2020},
}

@article{Wasielewski2023,
   author = {Michael R. Wasielewski},
   doi = {10.1063/PT.3.5196},
   issn = {0031-9228},
   issue = {3},
   journal = {Physics Today},
   month = {3},
   pages = {28-34},
   publisher = {American Institute of Physics},
   title = {Light-driven spin chemistry for quantum information science},
   volume = {76},
   url = {https://doi.org/10.1063/PT.3.5196 https://physicstoday.scitation.org/doi/10.1063/PT.3.5196},
   year = {2023},
}

@article{Quintes2023,
   author = {Theresia Quintes and Maximilian Mayländer and Sabine Richert},
   doi = {10.1038/s41570-022-00453-y},
   issn = {2397-3358},
   issue = {2},
   journal = {Nature Reviews Chemistry},
   month = {1},
   pages = {75-90},
   publisher = {Springer US},
   title = {Properties and applications of photoexcited chromophore–radical systems},
   volume = {7},
   url = {https://www.nature.com/articles/s41570-022-00453-y},
   year = {2023},
}

@article{Moro2022,
   author = {Fabrizio Moro and Massimo Moret and Alberto Ghirri and Andrés Granados del Águila and Yoshihiro Kubozono and Luca Beverina and Antonio Cassinese},
   doi = {10.1557/s43578-022-00536-y},
   isbn = {0123456789},
   issn = {20445326},
   issue = {6},
   journal = {Journal of Materials Research},
   pages = {1269-1279},
   publisher = {Springer International Publishing},
   title = {Room-temperature optically detected magnetic resonance of triplet excitons in a pentacene-doped picene single crystal},
   volume = {37},
   url = {https://doi.org/10.1557/s43578-022-00536-y},
   year = {2022},
}

@article{Gundel1989,
   author = {D. Gundel and J. Frick and J. Krzystek and H. Sixl and J. U. von Schütz and H. C. Wolf},
   doi = {10.1016/0301-0104(89)80030-8},
   issn = {03010104},
   issue = {3},
   journal = {Chemical Physics},
   pages = {363-372},
   title = {A quasi-neutral triplet state of TCNQ in phenazine/TCNQ and fluorene/TCNQ CT crystals},
   volume = {132},
   year = {1989},
}

@article{Clarke1976,
   author = {Richard H. Clarke and Harry A. Frank},
   doi = {10.1063/1.432781},
   issn = {00219606},
   issue = {1},
   journal = {The Journal of Chemical Physics},
   pages = {39-47},
   title = {Triplet state radiationless transitions in polycyclic hydrocarbons},
   volume = {65},
   year = {1976},
}

@article{Brauchle1979,
   author = {Chr Bräuchle and H Kabza and J Voitländer},
   journal = {Z. Naturforsch},
   pages = {6-12},
   title = {Optical and ODMR Investigations of the Lowest Excited Triplet State of Dinaphtho-(2'.3':1.2); (2".3":6.7)-pyrene},
   volume = {34a},
   year = {1979},
}

@article{Agostini1993,
   author = {Giancarlo Agostini and Carlo Corvaja and Giovanni Giacometti and Luigi Pasimeni},
   doi = {10.1016/0301-0104(93)80139-Z},
   issn = {03010104},
   issue = {2},
   journal = {Chemical Physics},
   month = {6},
   pages = {177-186},
   title = {Optical, zero-field ODMR and EPR studies of the triplet states from singlet fission in biphenyl-TCNQ and biphenyl-tetrafluoro-TCNQ charge-transfer crystals},
   volume = {173},
   url = {https://linkinghub.elsevier.com/retrieve/pii/030101049380139Z},
   year = {1993},
}

@article{Corvaja2005,
   author = {C. Corvaja and L. Franco and K. M. Salikhov and V. K. Voronkova},
   doi = {10.1007/BF03166755},
   issn = {0937-9347},
   issue = {3-4},
   journal = {Applied Magnetic Resonance},
   month = {9},
   pages = {181-193},
   title = {The first observation of electron spin polarization in the excited triplet states caused by the triplet-triplet annihilation},
   volume = {28},
   url = {http://link.springer.com/10.1007/BF03166755},
   year = {2005},
}

@article{Corvaja1992,
   author = {Carlo Corvaja and L. Franco and L. Pasimeni and A. Toffoletti},
   doi = {10.1007/BF03260112},
   issn = {16137507},
   issue = {5},
   journal = {Applied Magnetic Resonance},
   pages = {797-813},
   title = {Time resolved EPR of triplet excitons in Phenazine-TCNQ charge transfer crystal},
   volume = {3},
   year = {1992},
}

@article{Chowdhury2024,
   author = {Rituparno Chowdhury and Petri Murto and Naitik A. Panjwani and Yan Sun and Pratyush Ghosh and Yorrick Boeije and Chiara Delpiano Cordeiro and Vadim Derkach and Seung-Je Woo and Oliver Millington and Daniel G. Congrave and Yao Fu and Tarig B. E. Mustafa and Miguel Monteverde and Jesús Cerdá and Giacomo Londi and Jan Behrends and Akshay Rao and David Beljonne and Alexei Chepelianskii and Hugo Bronstein and Richard H. Friend},
   doi = {10.1038/s41557-025-01875-z},
   issn = {1755-4330},
   issue = {9},
   journal = {Nature Chemistry},
   month = {9},
   pages = {1410-1417},
   pmid = {40730683},
   publisher = {Nature Research},
   title = {Bright triplet and bright charge-separated singlet excitons in organic diradicals enable optical read-out and writing of spin states},
   volume = {17},
   url = {https://www.nature.com/articles/s41557-025-01875-z},
   year = {2025}
}

@article{Joshi2022,
   author = {Gajadhar Joshi and Ryan D. Dill and Karl J. Thorley and John E. Anthony and Obadiah G. Reid and Justin C. Johnson},
   doi = {10.1063/5.0103662},
   issn = {0021-9606},
   issue = {16},
   journal = {The Journal of Chemical Physics},
   month = {10},
   pages = {164702},
   pmid = {36319433},
   publisher = {AIP Publishing, LLC},
   title = {Optical readout of singlet fission biexcitons in a heteroacene with photoluminescence detected magnetic resonance},
   volume = {157},
   url = {https://doi.org/10.1063/5.0103662 https://aip.scitation.org/doi/10.1063/5.0103662 https://pubs.aip.org/jcp/article/157/16/164702/2842150/Optical-readout-of-singlet-fission-biexcitons-in-a},
   year = {2022},
}

@article{Sun2023,
   author = {Yan Sun and L. R. Weiss and V. Derkach and J. E. Anthony and M. Monteverde and A. D. Chepelianskii},
   doi = {10.1103/PhysRevB.108.155405},
   issn = {24699969},
   issue = {15},
   journal = {Physical Review B},
   month = {10},
   publisher = {American Physical Society},
   title = {Spin-dependent recombination mechanisms for quintet biexcitons generated through singlet fission},
   volume = {108},
   year = {2023},
}

@article{Grune2024,
   author = {Jeannine Grüne and Steph Montanaro and Thomas W. Bradbury and Ashish Sharma and Simon Dowland and Sebastian Gorgon and Oliver Millington and William K. Myers and Jan Behrends and Jenny Clark and Akshay Rao and Hugo Bronstein and Neil C. Greenham},
   month = {10},
   title = {High-Spin State Dynamics and Quintet-Mediated Emission in Intramolecular Singlet Fission},
   journal = {arXiv, 2410.07891},
   url = {http://arxiv.org/abs/2410.07891},
   year = {2024},
}

@article{barry2020,
archivePrefix = {arXiv},
arxivId = {1903.08176},
author = {Barry, John F. and Schloss, Jennifer M. and Bauch, Erik and Turner, Matthew J. and Hart, Connor A. and Pham, Linh M. and Walsworth, Ronald L.},
doi = {10.1103/RevModPhys.92.015004},
eprint = {1903.08176},
issn = {0034-6861},
journal = {Reviews of Modern Physics},
month = {Mar},
number = {1},
pages = {015004},
publisher = {American Physical Society},
title = {{Sensitivity optimization for NV-diamond magnetometry}},
url = {https://doi.org/10.1103/RevModPhys.92.015004 https://link.aps.org/doi/10.1103/RevModPhys.92.015004},
volume = {92},
year = {2020}
}

@article{Takahashi2008,
archivePrefix = {arXiv},
arxivId = {0804.1537},
author = {Takahashi, Susumu and Hanson, Ronald and van Tol, Johan and Sherwin, Mark S. and Awschalom, David D.},
doi = {10.1103/PhysRevLett.101.047601},
eprint = {0804.1537},
issn = {0031-9007},
journal = {Physical Review Letters},
month = {Jul},
number = {4},
pages = {047601},
title = {{Quenching Spin Decoherence in Diamond through Spin Bath Polarization}},
url = {https://link.aps.org/doi/10.1103/PhysRevLett.101.047601},
volume = {101},
year = {2008}
}

@article{Tsuji2024,
author = {Tsuji, Takeyuki and Sekiguchi, Takeharu and Iwasaki, Takayuki and Hatano, Mutsuko},
doi = {10.1002/qute.202300194},
issn = {2511-9044},
journal = {Advanced Quantum Technologies},
month = {Jan},
number = {1},
pages = {1--8},
title = {{Extending Spin Dephasing Time of Perfectly Aligned Nitrogen‐Vacancy Centers by Mitigating Stress Distribution on Highly Misoriented Chemical‐Vapor‐Deposition Diamond}},
url = {https://onlinelibrary.wiley.com/doi/10.1002/qute.202300194},
volume = {7},
year = {2024}
}

@article{Luo2022,
archivePrefix = {arXiv},
arxivId = {2111.07981},
author = {Luo, T. and Lindner, L. and Langer, J. and Cimalla, V. and Vidal, X. and Hahl, F. and Schreyvogel, C. and Onoda, S. and Ishii, S. and Ohshima, T. and Wang, D. and Simpson, D. A. and Johnson, B. C. and Capelli, M. and Blinder, R. and Jeske, J.},
doi = {10.1088/1367-2630/ac58b6},
eprint = {2111.07981},
issn = {1367-2630},
journal = {New Journal of Physics},
month = {Mar},
number = {3},
pages = {033030},
publisher = {IOP Publishing},
title = {{Creation of nitrogen-vacancy centers in chemical vapor deposition diamond for sensing applications}},
url = {https://iopscience.iop.org/article/10.1088/1367-2630/ac58b6},
volume = {24},
year = {2022}
}

@article{Zhang2023,
author = {Zhang, Xiaoran and Liu, Kang‐Yuan and Li, Fengjiao and Liu, Xiaobing and Duan, Shuai and Wang, Jia‐Ning and Liu, Gang‐Qin and Pan, Xin‐Yu and Chen, Xin and Zhang, Ping and Ma, Yanming and Chen, Changfeng},
doi = {10.1002/adfm.202309586},
issn = {1616-301X},
journal = {Advanced Functional Materials},
month = {Dec},
number = {52},
pages = {1--9},
title = {{Highly Coherent Nitrogen‐Vacancy Centers in Diamond via Rational High‐Pressure and High‐Temperature Synthesis and Treatment}},
url = {https://onlinelibrary.wiley.com/doi/10.1002/adfm.202309586},
volume = {33},
year = {2023}
}

@article{Zhang2017,
archivePrefix = {arXiv},
arxivId = {1709.09070},
author = {Zhang, Wenlong and Zhang, Jian and Wang, Junfeng and Feng, Fupan and Lin, Shengran and Lou, Liren and Zhu, Wei and Wang, Guanzhong},
doi = {10.1103/PhysRevB.96.235443},
eprint = {1709.09070},
issn = {2469-9950},
journal = {Physical Review B},
month = {Dec},
number = {23},
pages = {235443},
title = {{Depth-dependent decoherence caused by surface and external spins for NV centers in diamond}},
url = {https://link.aps.org/doi/10.1103/PhysRevB.96.235443},
volume = {96},
year = {2017}
}

@article{Feder2024,
   author = {Jacob S. Feder and Benjamin S. Soloway and Shreya Verma and Zhi Z. Geng and Shihao Wang and Bethel B. Kifle and Emmeline G. Riendeau and Yeghishe Tsaturyan and Leah R. Weiss and Mouzhe Xie and Jun Huang and Aaron Esser-Kahn and Laura Gagliardi and David D. Awschalom and Peter C. Maurer},
   doi = {10.1038/s41586-025-09417-w},
   issn = {0028-0836},
   issue = {8079},
   journal = {Nature},
   month = {9},
   pages = {73-79},
   publisher = {Nature Research},
   title = {A fluorescent-protein spin qubit},
   volume = {645},
   url = {https://www.nature.com/articles/s41586-025-09417-w},
   year = {2025}
}

@article{Karapatzakis2024,
archivePrefix = {arXiv},
arxivId = {2403.00521},
author = {Karapatzakis, Ioannis and Resch, Jeremias and Schrodin, Marcel and Fuchs, Philipp and Kieschnick, Michael and Heupel, Julia and Kussi, Luis and S{\"{u}}rgers, Christoph and Popov, Cyril and Meijer, Jan and Becher, Christoph and Wernsdorfer, Wolfgang and Hunger, David},
doi = {10.1103/PhysRevX.14.031036},
eprint = {2403.00521},
issn = {21603308},
journal = {Physical Review X},
number = {3},
pages = {31036},
publisher = {American Physical Society},
title = {{Microwave Control of the Tin-Vacancy Spin Qubit in Diamond with a Superconducting Waveguide}},
url = {https://doi.org/10.1103/PhysRevX.14.031036},
volume = {14},
year = {2024}
}

@article{Pingault2017,
archivePrefix = {arXiv},
arxivId = {1701.06848},
author = {Pingault, Benjamin and Jarausch, David-Dominik and Hepp, Christian and Klintberg, Lina and Becker, Jonas N. and Markham, Matthew and Becher, Christoph and Atat{\"{u}}re, Mete},
doi = {10.1038/ncomms15579},
eprint = {1701.06848},
issn = {2041-1723},
journal = {Nature Communications},
month = {May},
number = {1},
pages = {15579},
pmid = {28555618},
title = {{Coherent control of the silicon-vacancy spin in diamond}},
url = {https://www.nature.com/articles/ncomms15579},
volume = {8},
year = {2017}
}

@article{CChen2024,
author = {Chen, Chengke and Jiang, Bo and Hu, Xiaojun},
doi = {10.1080/26941112.2024.2332346},
issn = {2694-1112},
journal = {Functional Diamond},
month = {Dec},
number = {1},
pages = {2332346},
publisher = {Taylor & Francis},
title = {{Research Progress on Silicon Vacancy Color Centers in Diamond}},
url = {https://doi.org/10.1080/26941112.2024.2332346 https://www.tandfonline.com/doi/full/10.1080/26941112.2024.2332346},
volume = {4},
year = {2024}
}

@article{Gorlitz2022,
archivePrefix = {arXiv},
arxivId = {2110.05451},
author = {G{\"{o}}rlitz, Johannes and Herrmann, Dennis and Fuchs, Philipp and Iwasaki, Takayuki and Taniguchi, Takashi and Rogalla, Detlef and Hardeman, David and Colard, Pierre-Olivier and Markham, Matthew and Hatano, Mutsuko and Becher, Christoph},
doi = {10.1038/s41534-022-00552-0},
eprint = {2110.05451},
issn = {2056-6387},
journal = {npj Quantum Information},
month = {Apr},
number = {1},
pages = {45},
publisher = {Springer US},
title = {{Coherence of a charge stabilised tin-vacancy spin in diamond}},
url = {https://www.nature.com/articles/s41534-022-00552-0},
volume = {8},
year = {2022}
}

@incollection{Thiering2020,
author = {Thiering, Gergő and Gali, Adam},
booktitle = {Semiconductors and Semimetals},
doi = {10.1016/bs.semsem.2020.03.001},
isbn = {9780128202401},
issn = {00808784},
pages = {1--36},
title = {{Color centers in diamond for quantum applications}},
url = {https://linkinghub.elsevier.com/retrieve/pii/S0080878420300016},
volume = {103},
year = {2020},
publisher = {Elsevier}
}

@article{Balasubramanian2009,
author = {Balasubramanian, Gopalakrishnan and Neumann, Philipp and Twitchen, Daniel and Markham, Matthew and Kolesov, Roman and Mizuochi, Norikazu and Isoya, Junichi and Achard, Jocelyn and Beck, Johannes and Tissler, Julia and Jacques, Vincent and Hemmer, Philip R. and Jelezko, Fedor and Wrachtrup, J{\"{o}}rg},
doi = {10.1038/nmat2420},
issn = {1476-1122},
journal = {Nature Materials},
month = {May},
number = {5},
pages = {383--387},
pmid = {19349970},
publisher = {Nature Publishing Group},
title = {{Ultralong spin coherence time in isotopically engineered diamond}},
url = {https://www.nature.com/articles/nmat2420},
volume = {8},
year = {2009}
}

@article{Morris2024,
   author = {I.M. Morris and T. Lühmann and K. Klink and L. Crooks and D. Hardeman and D.J. Twitchen and S. Pezzagna and J. Meijer and S.S. Nicley and J.N. Becker},
   doi = {10.1103/wzjx-z2h5},
   issn = {0031-9007},
   issue = {4},
   journal = {Physical Review Letters},
   month = {7},
   pages = {043602},
   title = {Lifetime-Limited and Tunable Emission from Single Charge-Stabilized Nickel Vacancy Centers in Diamond},
   volume = {135},
   url = {https://link.aps.org/doi/10.1103/wzjx-z2h5},
   year = {2025}
}

@article{Thiering2021,
author = {Thiering, Gergő and Gali, Adam},
doi = {10.1103/PhysRevResearch.3.043052},
issn = {2643-1564},
journal = {Physical Review Research},
month = {Oct},
number = {4},
pages = {043052},
title = {{Magneto-optical spectra of the split nickel-vacancy defect in diamond}},
url = {https://link.aps.org/doi/10.1103/PhysRevResearch.3.043052},
volume = {3},
year = {2021}
}

@article{Waltrich2023,
author = {Waltrich, Richard and Klotz, Marco and Agafonov, Viatcheslav N. and Kubanek, Alexander},
doi = {10.1515/nanoph-2023-0379},
isbn = {0000000286084},
issn = {2192-8614},
journal = {Nanophotonics},
month = {Sep},
number = {18},
pages = {3663--3669},
title = {{Two-photon interference from silicon-vacancy centers in remote nanodiamonds}},
url = {https://doi.org/10.1515/nanoph-2023-0379 https://www.degruyter.com/document/doi/10.1515/nanoph-2023-0379/html},
volume = {12},
year = {2023}
}

@article{Bradac2019,
archivePrefix = {arXiv},
arxivId = {1906.10992},
author = {Bradac, Carlo and Gao, Weibo and Forneris, Jacopo and Trusheim, Matthew E. and Aharonovich, Igor},
doi = {10.1038/s41467-019-13332-w},
eprint = {1906.10992},
issn = {2041-1723},
journal = {Nature Communications},
month = {Dec},
number = {1},
pages = {5625},
pmid = {31819050},
publisher = {Springer US},
title = {{Quantum nanophotonics with group IV defects in diamond}},
url = {http://dx.doi.org/10.1038/s41467-019-13332-w https://www.nature.com/articles/s41467-019-13332-w},
volume = {10},
year = {2019}
}

@article{Green2017,
archivePrefix = {arXiv},
arxivId = {1705.10205},
author = {Green, B. L. and Mottishaw, S. and Breeze, B. G. and Edmonds, A. M. and D'Haenens-Johansson, U. F. S. and Doherty, M. W. and Williams, S. D. and Twitchen, D. J. and Newton, M. E.},
doi = {10.1103/PhysRevLett.119.096402},
eprint = {1705.10205},
issn = {0031-9007},
journal = {Physical Review Letters},
month = {Aug},
number = {9},
pages = {096402},
pmid = {28949565},
title = {{Neutral Silicon-Vacancy Center in Diamond: Spin Polarization and Lifetimes}},
url = {https://link.aps.org/doi/10.1103/PhysRevLett.119.096402},
volume = {119},
year = {2017}
}

@article{Chen2016,
  title = {Laser writing of coherent colour centres in diamond},
  volume = {11},
  ISSN = {1749-4893},
  url = {http://dx.doi.org/10.1038/nphoton.2016.234},
  DOI = {10.1038/nphoton.2016.234},
  number = {2},
  journal = {Nature Photonics},
  publisher = {Springer Science and Business Media LLC},
  author = {Chen,  Yu-Chen and Salter,  Patrick S. and Knauer,  Sebastian and Weng,  Laiyi and Frangeskou,  Angelo C. and Stephen,  Colin J. and Ishmael,  Shazeaa N. and Dolan,  Philip R. and Johnson,  Sam and Green,  Ben L. and Morley,  Gavin W. and Newton,  Mark E. and Rarity,  John G. and Booth,  Martin J. and Smith,  Jason M.},
  year = {2016},
  month = dec,
  pages = {77–80}
}

@article{Iwasaki2015,
author = {Iwasaki, Takayuki and Ishibashi, Fumitaka and Miyamoto, Yoshiyuki and Doi, Yuki and Kobayashi, Satoshi and Miyazaki, Takehide and Tahara, Kosuke and Jahnke, Kay D. and Rogers, Lachlan J. and Naydenov, Boris and Jelezko, Fedor and Yamasaki, Satoshi and Nagamachi, Shinji and Inubushi, Toshiro and Mizuochi, Norikazu and Hatano, Mutsuko},
doi = {10.1038/srep12882},
issn = {2045-2322},
journal = {Scientific Reports},
month = {Aug},
number = {1},
pages = {12882},
publisher = {Nature Publishing Group},
title = {{Germanium-Vacancy Single Color Centers in Diamond}},
url = {https://www.nature.com/articles/srep12882},
volume = {5},
year = {2015}
}

@article{Senkalla2024,
archivePrefix = {arXiv},
arxivId = {2308.09666},
author = {Senkalla, Katharina and Genov, Genko and Metsch, Mathias H. and Siyushev, Petr and Jelezko, Fedor},
doi = {10.1103/PhysRevLett.132.026901},
eprint = {2308.09666},
issn = {0031-9007},
journal = {Physical Review Letters},
month = {Jan},
number = {2},
pages = {026901},
pmid = {38277597},
publisher = {American Physical Society},
title = {{Germanium Vacancy in Diamond Quantum Memory Exceeding 20 ms}},
url = {https://doi.org/10.1103/PhysRevLett.132.026901 https://link.aps.org/doi/10.1103/PhysRevLett.132.026901},
volume = {132},
year = {2024}
}

@article{Stanwix2010,
archivePrefix = {arXiv},
arxivId = {1006.4219},
author = {Stanwix, P. L. and Pham, L. M. and Maze, J. R. and {Le Sage}, D. and Yeung, T. K. and Cappellaro, P. and Hemmer, P. R. and Yacoby, A. and Lukin, M. D. and Walsworth, R. L.},
doi = {10.1103/PhysRevB.82.201201},
eprint = {1006.4219},
issn = {1098-0121},
journal = {Physical Review B},
month = {Nov},
number = {20},
pages = {201201},
title = {{Coherence of nitrogen-vacancy electronic spin ensembles in diamond}},
url = {https://link.aps.org/doi/10.1103/PhysRevB.82.201201},
volume = {82},
year = {2010}
}

@article{Umeda2022,
author = {Umeda, T. and Watanabe, K. and Hara, H. and Sumiya, H. and Onoda, S. and Uedono, A. and Chuprina, I. and Siyushev, P. and Jelezko, F. and Wrachtrup, J. and Isoya, J.},
doi = {10.1103/PhysRevB.105.165201},
issn = {2469-9950},
journal = {Physical Review B},
month = {Apr},
number = {16},
pages = {165201},
publisher = {American Physical Society},
title = {{Negatively charged boron vacancy center in diamond}},
url = {https://link.aps.org/doi/10.1103/PhysRevB.105.165201},
volume = {105},
year = {2022}
}

@article{Komarovskikh2014,
author = {Komarovskikh, Andrey and Nadolinny, Vladimir and Palyanov, Yuri and Kupriyanov, Igor and Sokol, Alexander},
doi = {10.1002/pssa.201431163},
issn = {1862-6300},
journal = {Physica status solidi (a)},
month = {Oct},
number = {10},
pages = {2274--2278},
title = {{EPR study of the hydrogen center in HPHT diamonds grown in carbonate medium}},
url = {https://onlinelibrary.wiley.com/doi/10.1002/pssa.201431163},
volume = {211},
year = {2014}
}

@phdthesis{Hartland2014,
author = {Hartland, Christopher Brett},
school = {University of Warwick},
booktitle = {School University of Warwick},
pages = {1--231},
title = {{A Study of Point Defects in CVD Diamond Using Electron Paramagnetic Resonance and Optical Spectroscopy}},
year = {2014}
}

@article{Trusheim2019,
archivePrefix = {arXiv},
arxivId = {1805.12202},
author = {Trusheim, Matthew E. and Wan, Noel H. and Chen, Kevin C. and Ciccarino, Christopher J. and Flick, Johannes and Sundararaman, Ravishankar and Malladi, Girish and Bersin, Eric and Walsh, Michael and Lienhard, Benjamin and Bakhru, Hassaram and Narang, Prineha and Englund, Dirk},
doi = {10.1103/PhysRevB.99.075430},
eprint = {1805.12202},
issn = {2469-9950},
journal = {Physical Review B},
month = {Feb},
number = {7},
pages = {075430},
publisher = {American Physical Society},
title = {{Lead-related quantum emitters in diamond}},
url = {https://link.aps.org/doi/10.1103/PhysRevB.99.075430},
volume = {99},
year = {2019}
}

@article{Wang2024,
archivePrefix = {arXiv},
arxivId = {2308.00995},
author = {Wang, Peng and Kazak, Lev and Senkalla, Katharina and Siyushev, Petr and Abe, Ryotaro and Taniguchi, Takashi and Onoda, Shinobu and Kato, Hiromitsu and Makino, Toshiharu and Hatano, Mutsuko and Jelezko, Fedor and Iwasaki, Takayuki},
doi = {10.1103/PhysRevLett.132.073601},
eprint = {2308.00995},
issn = {0031-9007},
journal = {Physical Review Letters},
month = {Feb},
number = {7},
pages = {073601},
pmid = {38427893},
publisher = {American Physical Society},
title = {{Transform-Limited Photon Emission from a Lead-Vacancy Center in Diamond above 10 K}},
url = {https://doi.org/10.1103/PhysRevLett.132.073601 https://link.aps.org/doi/10.1103/PhysRevLett.132.073601},
volume = {132},
year = {2024}
}

@article{Trusheim2020,
archivePrefix = {arXiv},
arxivId = {1811.07777},
author = {Trusheim, Matthew E. and Pingault, Benjamin and Wan, Noel H. and G{\"{u}}ndoğan, Mustafa and {De Santis}, Lorenzo and Debroux, Romain and Gangloff, Dorian and Purser, Carola and Chen, Kevin C. and Walsh, Michael and Rose, Joshua J. and Becker, Jonas N. and Lienhard, Benjamin and Bersin, Eric and Paradeisanos, Ioannis and Wang, Gang and Lyzwa, Dominika and Montblanch, Alejandro R.P. and Malladi, Girish and Bakhru, Hassaram and Ferrari, Andrea C. and Walmsley, Ian A. and Atat{\"{u}}re, Mete and Englund, Dirk},
doi = {10.1103/PhysRevLett.124.023602},
eprint = {1811.07777},
issn = {0031-9007},
journal = {Physical Review Letters},
month = {Jan},
number = {2},
pages = {023602},
pmid = {32004012},
title = {{Transform-Limited Photons From a Coherent Tin-Vacancy Spin in Diamond}},
url = {https://link.aps.org/doi/10.1103/PhysRevLett.124.023602},
volume = {124},
year = {2020}
}

@article{Meesala2018,
archivePrefix = {arXiv},
arxivId = {1801.09833},
author = {Meesala, Srujan and Sohn, Young-Ik and Pingault, Benjamin and Shao, Linbo and Atikian, Haig A. and Holzgrafe, Jeffrey and G{\"{u}}ndoğan, Mustafa and Stavrakas, Camille and Sipahigil, Alp and Chia, Cleaven and Evans, Ruffin and Burek, Michael J. and Zhang, Mian and Wu, Lue and Pacheco, Jose L. and Abraham, John and Bielejec, Edward and Lukin, Mikhail D. and Atat{\"{u}}re, Mete and Lon{\v{c}}ar, Marko},
doi = {10.1103/PhysRevB.97.205444},
eprint = {1801.09833},
issn = {2469-9950},
journal = {Physical Review B},
month = {May},
number = {20},
pages = {205444},
title = {{Strain engineering of the silicon-vacancy center in diamond}},
url = {https://link.aps.org/doi/10.1103/PhysRevB.97.205444},
volume = {97},
year = {2018}
}

@article{Sohn2018,
author = {Sohn, Young-Ik and Meesala, Srujan and Pingault, Benjamin and Atikian, Haig A. and Holzgrafe, Jeffrey and G{\"{u}}ndoğan, Mustafa and Stavrakas, Camille and Stanley, Megan J. and Sipahigil, Alp and Choi, Joonhee and Zhang, Mian and Pacheco, Jose L. and Abraham, John and Bielejec, Edward and Lukin, Mikhail D. and Atat{\"{u}}re, Mete and Lon{\v{c}}ar, Marko},
doi = {10.1038/s41467-018-04340-3},
issn = {2041-1723},
journal = {Nature Communications},
month = {May},
number = {1},
pages = {2012},
pmid = {29789553},
title = {{Controlling the coherence of a diamond spin qubit through its strain environment}},
url = {https://www.nature.com/articles/s41467-018-04340-3},
volume = {9},
year = {2018}
}

@article{Gaita-Arino2019,
author = {Gaita-Ari{\~{n}}o, A. and Luis, F. and Hill, S. and Coronado, E.},
doi = {10.1038/s41557-019-0232-y},
issn = {1755-4330},
journal = {Nature Chemistry},
month = {Apr},
number = {4},
pages = {301--309},
pmid = {30903036},
publisher = {Springer US},
title = {{Molecular spins for quantum computation}},
url = {http://dx.doi.org/10.1038/s41557-019-0232-y https://www.nature.com/articles/s41557-019-0232-y},
volume = {11},
year = {2019}
}

@incollection{Aromi2019,
author = {Arom{\'{i}}, Guillem and Roubeau, Olivier},
booktitle = {Handbook on the Physics and Chemistry of Rare Earths},
doi = {10.1016/bs.hpcre.2019.07.002},
publisher = {Elsevier}, 
isbn = {9780444642998},
issn = {01681273},
pages = {1--54},
title = {{Lanthanide molecules for spin-based quantum technologies}},
url = {https://linkinghub.elsevier.com/retrieve/pii/S0168127319300042},
volume = {56},
year = {2019}
}

@book{Sorace2015,
author = {Sorace, Lorenzo and Gatteschi, Dante and Clemente-Juan, Juan M. and Coronado, Eugenio and Gaita-Ari{\~{n}}o, Alejandro and Tang, Jinkui and Zhang, Peng and Sessoli, Roberta and Bernot, Kevin and {Kasper S. Pedersen}, Daniel N. Woodruff and Bendix, Jesper and Cl{\'{e}}rac, Rodolphe and Ungur, Liviu and Chibotaru, Liviu F. and Arom{\'{i}}, Guillem and Luis, Fernando and Roubeau, Olivier and Lan, Yanhua and Klyatskaya, Svetlana and Ruben, Mario and Sharples, Joseph W. and Collison, David and Liddle, Stephen T. and van Slageren, Joris},
doi = {10.1002/9783527673476},
isbn = {9783527335268},
month = {Mar},
publisher = {Wiley},
title = {{Lanthanides and Actinides in Molecular Magnetism}},
url = {https://onlinelibrary.wiley.com/doi/book/10.1002/9783527673476},
year = {2015}
}

@article{Zhong2019,
author = {Zhong, Manjin and Ahlefeldt, Rose L. and Sellars, Matthew J.},
doi = {10.1088/1367-2630/ab0cb7},
issn = {1367-2630},
journal = {New Journal of Physics},
month = {Mar},
number = {3},
pages = {033019},
publisher = {IOP Publishing},
title = {{Quantum information processing using frozen core Y$^{3+}$ spins in Eu$^{3+}$:Y$_2$SiO$_5$}},
url = {https://iopscience.iop.org/article/10.1088/1367-2630/ab0cb7},
volume = {21},
year = {2019}
}

@article{Bruzewicz2019,
archivePrefix = {arXiv},
arxivId = {1904.04178},
author = {Bruzewicz, Colin D. and Chiaverini, John and McConnell, Robert and Sage, Jeremy M.},
doi = {10.1063/1.5088164},
eprint = {1904.04178},
issn = {1931-9401},
journal = {Applied Physics Reviews},
month = {Jun},
number = {2},
publisher = {AIP Publishing LLC},
title = {{Trapped-ion quantum computing: Progress and challenges}},
url = {https://pubs.aip.org/apr/article/6/2/021314/570103/Trapped-ion-quantum-computing-Progress-and},
volume = {6},
year = {2019}
}

@article{Brown2021,
archivePrefix = {arXiv},
arxivId = {2009.00568},
author = {Brown, Kenneth R. and Chiaverini, John and Sage, Jeremy M. and H{\"{a}}ffner, Hartmut},
doi = {10.1038/s41578-021-00292-1},
eprint = {2009.00568},
isbn = {0123456789},
issn = {20588437},
journal = {Nature Reviews Materials},
number = {10},
pages = {892--905},
publisher = {Springer US},
title = {{Materials challenges for trapped-ion quantum computers}},
url = {http://dx.doi.org/10.1038/s41578-021-00292-1},
volume = {6},
year = {2021}
}

@article{Moses2023,
archivePrefix = {arXiv},
arxivId = {2305.03828},
author = {Moses, S. A. and Baldwin, C. H. and Allman, M. S. and Ancona, R. and Ascarrunz, L. and Barnes, C. and Bartolotta, J. and Bjork, B. and Blanchard, P. and Bohn, M. and Bohnet, J. G. and Brown, N. C. and Burdick, N. Q. and Burton, W. C. and Campbell, S. L. and Campora, J. P. and Carron, C. and Chambers, J. and Chan, J. W. and Chen, Y. H. and Chernoguzov, A. and Chertkov, E. and Colina, J. and Curtis, J. P. and Daniel, R. and DeCross, M. and Deen, D. and Delaney, C. and Dreiling, J. M. and Ertsgaard, C. T. and Esposito, J. and Estey, B. and Fabrikant, M. and Figgatt, C. and Foltz, C. and Foss-Feig, M. and Francois, D. and Gaebler, J. P. and Gatterman, T. M. and Gilbreth, C. N. and Giles, J. and Glynn, E. and Hall, A. and Hankin, A. M. and Hansen, A. and Hayes, D. and Higashi, B. and Hoffman, I. M. and Horning, B. and Hout, J. J. and Jacobs, R. and Johansen, J. and Jones, L. and Karcz, J. and Klein, T. and Lauria, P. and Lee, P. and Liefer, D. and Lu, S. T. and Lucchetti, D. and Lytle, C. and Malm, A. and Matheny, M. and Mathewson, B. and Mayer, K. and Miller, D. B. and Mills, M. and Neyenhuis, B. and Nugent, L. and Olson, S. and Parks, J. and Price, G. N. and Price, Z. and Pugh, M. and Ransford, A. and Reed, A. P. and Roman, C. and Rowe, M. and Ryan-Anderson, C. and Sanders, S. and Sedlacek, J. and Shevchuk, P. and Siegfried, P. and Skripka, T. and Spaun, B. and Sprenkle, R. T. and Stutz, R. P. and Swallows, M. and Tobey, R. I. and Tran, A. and Tran, T. and Vogt, E. and Volin, C. and Walker, J. and Zolot, A. M. and Pino, J. M.},
doi = {10.1103/PhysRevX.13.041052},
eprint = {2305.03828},
issn = {2160-3308},
journal = {Physical Review X},
month = {Dec},
number = {4},
pages = {041052},
publisher = {American Physical Society},
title = {{A Race-Track Trapped-Ion Quantum Processor}},
url = {https://doi.org/10.1103/PhysRevX.13.041052 https://link.aps.org/doi/10.1103/PhysRevX.13.041052},
volume = {13},
year = {2023}
}

@article{Fraval2005,
author = {Fraval, E. and Sellars, M. J. and Longdell, J. J.},
doi = {10.1103/PhysRevLett.95.030506},
issn = {00319007},
journal = {Physical Review Letters},
number = {3},
pages = {8--11},
title = {{Dynamic decoherence control of a solid-state nuclear-quadrupole qubit}},
volume = {95},
year = {2005}
}

@article{Fraval2004,
author = {Fraval, E. and Sellars, M. J. and Longdell, J. J.},
doi = {10.1103/PhysRevLett.92.077601},
issn = {0031-9007},
journal = {Physical Review Letters},
month = {Feb},
number = {7},
pages = {077601},
title = {{Method of Extending Hyperfine Coherence Times in Pr$^{3+}$:Y$_2$SiO$_5$}},
url = {https://link.aps.org/doi/10.1103/PhysRevLett.92.077601},
volume = {92},
year = {2004}
}

@article{Yano1991,
author = {Yano, Ryuzi and Mitsunaga, Masaharu and Uesugi, Naoshi},
doi = {10.1364/OL.16.001884},
issn = {0146-9592},
journal = {Optics Letters},
month = {Dec},
number = {23},
pages = {1884},
title = {{Ultralong optical dephasing time in Eu$^{3+}$:Y$_2$SiO$_5$}},
url = {https://opg.optica.org/abstract.cfm?URI=ol-16-23-1884},
volume = {16},
year = {1991}
}

@article{McAuslan2012,
archivePrefix = {arXiv},
arxivId = {1201.4610},
author = {McAuslan, D. L. and Bartholomew, J. G. and Sellars, M. J. and Longdell, J. J.},
doi = {10.1103/PhysRevA.85.032339},
eprint = {1201.4610},
issn = {1050-2947},
journal = {Physical Review A},
month = {Mar},
number = {3},
pages = {032339},
title = {{Reducing decoherence in optical and spin transitions in rare-earth-metal-ion–doped materials}},
url = {https://link.aps.org/doi/10.1103/PhysRevA.85.032339},
volume = {85},
year = {2012}
}

@article{Zhong2015,
author = {Zhong, Manjin and Hedges, Morgan P. and Ahlefeldt, Rose L. and Bartholomew, John G. and Beavan, Sarah E. and Wittig, Sven M. and Longdell, Jevon J. and Sellars, Matthew J.},
doi = {10.1038/nature14025},
issn = {0028-0836},
journal = {Nature},
month = {Jan},
number = {7533},
pages = {177--180},
publisher = {Nature Publishing Group},
title = {{Optically addressable nuclear spins in a solid with a six-hour coherence time}},
url = {https://www.nature.com/articles/nature14025},
volume = {517},
year = {2015}
}

@article{Jobbitt2021,
archivePrefix = {arXiv},
arxivId = {2104.11415},
author = {Jobbitt, N. L. and Wells, J.-P. R. and Reid, M. F. and Longdell, J. J.},
doi = {10.1103/PhysRevB.103.205114},
eprint = {2104.11415},
issn = {2469-9950},
journal = {Physical Review B},
month = {May},
number = {20},
pages = {205114},
publisher = {American Physical Society},
title = {{Raman heterodyne determination of the magnetic anisotropy for the ground and optically excited states of Y$_2$SiO$_5$ doped with Sm$^{3+}$}},
url = {https://link.aps.org/doi/10.1103/PhysRevB.103.205114},
volume = {103},
year = {2021}
}

@article{Jobbitt2022,
author = {Jobbitt, N. L. and Wells, J-P R and Reid, M. F.},
doi = {10.1088/1361-648X/ac711e},
issn = {0953-8984},
journal = {Journal of Physics: Condensed Matter},
month = {Aug},
number = {32},
pages = {325502},
pmid = {35584691},
publisher = {IOP Publishing},
title = {{Zeeman and laser site selective spectroscopy of C 1 point group symmetry Sm$^{3+}$ centres in Y$_2$SiO$_5$: a parametrized crystal-field analysis for the 4f 5 configuration}},
url = {https://iopscience.iop.org/article/10.1088/1361-648X/ac711e},
volume = {34},
year = {2022}
}

@article{Heinze2014,
author = {Heinze, G. and Hubrich, C. and Halfmann, T.},
doi = {10.1103/PhysRevA.89.053825},
issn = {1050-2947},
journal = {Physical Review A},
month = {May},
number = {5},
pages = {053825},
title = {{Hyperfine characterization and spin coherence lifetime extension in Pr$^{3+}$:La$_2$(WO$_4$)$_3$}},
url = {https://link.aps.org/doi/10.1103/PhysRevA.89.053825},
volume = {89},
year = {2014}
}

@article{Ortu2018,
archivePrefix = {arXiv},
arxivId = {1712.08615},
author = {Ortu, Antonio and Tiranov, Alexey and Welinski, Sacha and Fr{\"{o}}wis, Florian and Gisin, Nicolas and Ferrier, Alban and Goldner, Philippe and Afzelius, Mikael},
doi = {10.1038/s41563-018-0138-x},
eprint = {1712.08615},
issn = {1476-1122},
journal = {Nature Materials},
month = {Aug},
number = {8},
pages = {671--675},
pmid = {30042512},
publisher = {Springer US},
title = {{Simultaneous coherence enhancement of optical and microwave transitions in solid-state electronic spins}},
url = {http://dx.doi.org/10.1038/s41563-018-0138-x https://www.nature.com/articles/s41563-018-0138-x},
volume = {17},
year = {2018}
}

@article{Lovric2011,
archivePrefix = {arXiv},
arxivId = {1107.2274},
author = {Lovri{\'{c}}, Marko and Glasenapp, Philipp and Suter, Dieter and Tumino, Biagio and Ferrier, Alban and Goldner, Philippe and Sabooni, Mahmood and Rippe, Lars and Kr{\"{o}}ll, Stefan},
doi = {10.1103/PhysRevB.84.104417},
eprint = {1107.2274},
issn = {1098-0121},
journal = {Physical Review B},
month = {Sep},
number = {10},
pages = {104417},
title = {{Hyperfine characterization and spin coherence lifetime extension in Pr$^{3+}$:La$_2$(WO$_4$)$_3$}},
url = {https://link.aps.org/doi/10.1103/PhysRevB.84.104417},
volume = {84},
year = {2011}
}

@article{Rancic2018,
archivePrefix = {arXiv},
arxivId = {1611.04315},
author = {Ran{\v{c}}i{\'{c}}, Milo{\v{s}} and Hedges, Morgan P. and Ahlefeldt, Rose L. and Sellars, Matthew J.},
doi = {10.1038/nphys4254},
eprint = {1611.04315},
issn = {1745-2473},
journal = {Nature Physics},
month = {Jan},
number = {1},
pages = {50--54},
title = {{Coherence time of over a second in a telecom-compatible quantum memory storage material}},
url = {https://www.nature.com/articles/nphys4254},
volume = {14},
year = {2018}
}

@article{Shoudu1999,
author = {Shoudu, Zhang and Siting, Wang and Xingda, Shen and Haobing, Wang and Heyu, Zhong and Shunxing, Zhang and Jun, Xu},
doi = {10.1016/S0022-0248(98)00553-3},
issn = {00220248},
journal = {Journal of Crystal Growth},
month = {Mar},
number = {4},
pages = {901-904},
title = {{Czochralski growth of rare-earth orthosilicates–Y$_2$SiO$_5$ single crystals}},
url = {https://linkinghub.elsevier.com/retrieve/pii/S0022024898005533},
volume = {197},
year = {1999}
}

@article{Bottger2016,
author = {B{\"{o}}ttger, Thomas and Thiel, C.W. and Sun, Y. and Macfarlane, R.M. and Cone, R.L.},
doi = {10.1016/j.jlumin.2014.11.014},
issn = {00222313},
journal = {Journal of Luminescence},
month = {Jan},
pages = {466--471},
title = {{Decoherence and absorption of Er$^{3+}$:KTiOPO$_4$ (KTP) at 1.5 $\mu$m}},
url = {https://linkinghub.elsevier.com/retrieve/pii/S0022231314006553},
volume = {169},
year = {2016}
}

@article{Lim2018,
archivePrefix = {arXiv},
arxivId = {1712.00435},
author = {Lim, Hee-Jin and Welinski, Sacha and Ferrier, Alban and Goldner, Philippe and Morton, J. J. L.},
doi = {10.1103/PhysRevB.97.064409},
eprint = {1712.00435},
issn = {2469-9950},
journal = {Physical Review B},
month = {Feb},
number = {6},
pages = {064409},
publisher = {American Physical Society},
title = {{Coherent spin dynamics of ytterbium ions in yttrium orthosilicate}},
url = {https://link.aps.org/doi/10.1103/PhysRevB.97.064409},
volume = {97},
year = {2018}
}

@article{Welinski2020,
archivePrefix = {arXiv},
arxivId = {1910.07907},
author = {Welinski, Sacha and Tiranov, Alexey and Businger, Moritz and Ferrier, Alban and Afzelius, Mikael and Goldner, Philippe},
doi = {10.1103/PhysRevX.10.031060},
eprint = {1910.07907},
issn = {2160-3308},
journal = {Physical Review X},
month = {Sep},
number = {3},
pages = {031060},
publisher = {American Physical Society},
title = {{Coherence Time Extension by Large-Scale Optical Spin Polarization in a Rare-Earth Doped Crystal}},
url = {https://doi.org/10.1103/PhysRevX.10.031060 https://link.aps.org/doi/10.1103/PhysRevX.10.031060},
volume = {10},
year = {2020}
}

@article{Laorenza2021,
author = {Laorenza, Daniel W. and Kairalapova, Arailym and Bayliss, Sam L. and Goldzak, Tamar and Greene, Samuel M. and Weiss, Leah R. and Deb, Pratiti and Mintun, Peter J. and Collins, Kelsey A. and Awschalom, David D. and Berkelbach, Timothy C. and Freedman, Danna E.},
doi = {10.1021/jacs.1c10145},
issn = {15205126},
journal = {Journal of the American Chemical Society},
number = {50},
pages = {21350--21363},
pmid = {34817994},
title = {{Tunable Cr$^{4+}$ Molecular Color Centers}},
volume = {143},
year = {2021}
}

@article{Bayliss2022,
archivePrefix = {arXiv},
arxivId = {2204.00168},
author = {Bayliss, S. L. and Deb, P. and Laorenza, D. W. and Onizhuk, M. and Galli, G. and Freedman, D. E. and Awschalom, D. D.},
doi = {10.1103/physrevx.12.031028},
eprint = {2204.00168},
issn = {21603308},
journal = {Physical Review X},
number = {3},
pages = {31028},
publisher = {American Physical Society},
title = {{Enhancing Spin Coherence in Optically Addressable Molecular Qubits through Host-Matrix Control}},
url = {https://doi.org/10.1103/PhysRevX.12.031028},
volume = {12},
year = {2022}
}

@article{Bayliss2020,
archivePrefix = {arXiv},
arxivId = {2004.07998},
author = {Bayliss, S. L. and Laorenza, D. W. and Mintun, P. J. and Kovos, B. D. and Freedman, D. E. and Awschalom, D. D.},
doi = {10.1126/science.abb9352},
eprint = {2004.07998},
issn = {0036-8075},
journal = {Science},
month = {Dec},
number = {6522},
pages = {1309--1312},
pmid = {33184235},
title = {{Optically addressable molecular spins for quantum information processing}},
url = {https://www.science.org/doi/10.1126/science.abb9352},
volume = {370},
year = {2020}
}

@article{Fataftah2018,
author = {Fataftah, Majed S. and Freedman, Danna E.},
doi = {10.1039/C8CC07939K},
issn = {1364548X},
journal = {Chemical Communications},
number = {98},
pages = {13773--13781},
publisher = {Royal Society of Chemistry},
title = {{Progress towards creating optically addressable molecular qubits}},
volume = {54},
year = {2018}
}

@article{Yu2021,
author = {Yu, Chung-Jui and von Kugelgen, Stephen and Laorenza, Daniel W. and Freedman, Danna E.},
doi = {10.1021/acscentsci.0c00737},
issn = {2374-7943},
journal = {ACS Central Science},
month = {May},
number = {5},
pages = {712--723},
title = {{A Molecular Approach to Quantum Sensing}},
url = {https://pubs.acs.org/doi/10.1021/acscentsci.0c00737},
volume = {7},
year = {2021}
}

@article{Wrachtrup1993,
author = {Wrachtrup, J. and von Borczyskowski, C. and Bernard, J. and Orrit, M. and Brown, R.},
doi = {10.1103/PhysRevLett.71.3565},
issn = {0031-9007},
journal = {Physical Review Letters},
month = {Nov},
number = {21},
pages = {3565--3568},
title = {{Optically detected spin coherence of single molecules}},
url = {https://link.aps.org/doi/10.1103/PhysRevLett.71.3565},
volume = {71},
year = {1993}
}

@article{Wrachtrup1993b,
author = {Wrachtrup, J. and von Borczyskowski, C. and Bernard, J. and Orrit, Michel and Brown, R.},
doi = {10.1038/363244a0},
issn = {0028-0836},
journal = {Nature},
month = {May},
number = {6426},
pages = {244--245},
title = {{Optical detection of magnetic resonance in a single molecule}},
url = {https://www.nature.com/articles/363244a0},
volume = {363},
year = {1993}
}

@inproceedings{Kilin1998,
author = {Kilin, Sergei Y. and Nizovtsev, Alexander P. and Berman, Paul R. and Wrachtrup, Joerg and von Borczyskowski, Christian},
booktitle = {11th International Vavilov Conference on Nonlinear Optics},
doi = {10.1117/12.328223},
issn = {1996756X},
pages = {98},
title = {{Fluorescence-detected coherent phenomena on single triplet-state molecules}},
url = {http://spiedl.org/terms},
volume = {3485},
year = {1998}
}

@article{Lew2024,
author = {Lew, C. T.-K. and Sewani, V. K. and Iwamoto, N. and Ohshima, T. and McCallum, J. C. and Johnson, B. C.},
doi = {10.1103/PhysRevLett.132.146902},
issn = {0031-9007},
journal = {Physical Review Letters},
month = {Apr},
number = {14},
pages = {146902},
pmid = {38640398},
publisher = {American Physical Society},
title = {{All-Electrical Readout of Coherently Controlled Spins in Silicon Carbide}},
url = {https://doi.org/10.1103/PhysRevLett.132.146902 https://link.aps.org/doi/10.1103/PhysRevLett.132.146902},
volume = {132},
year = {2024}
}

@article{Ou2024,
author = {Ou, Haiyan},
doi = {10.1038/s41377-024-01515-0},
issn = {2047-7538},
journal = {Light: Science \& Applications},
month = {Aug},
number = {1},
pages = {219},
publisher = {Springer US},
title = {{Silicon carbide, the next-generation integrated platform for quantum technology}},
url = {http://dx.doi.org/10.1038/s41377-024-01515-0 https://www.nature.com/articles/s41377-024-01515-0},
volume = {13},
year = {2024}
}

@article{Lee2021a,
archivePrefix = {arXiv},
arxivId = {2109.06420},
author = {Lee, Elizabeth M. Y. and Yu, Alvin and de Pablo, Juan J. and Galli, Giulia},
doi = {10.1038/s41467-021-26419-0},
eprint = {2109.06420},
issn = {2041-1723},
journal = {Nature Communications},
month = {Nov},
number = {1},
pages = {6325},
pmid = {34732705},
publisher = {Springer US},
title = {{Stability and molecular pathways to the formation of spin defects in silicon carbide}},
url = {https://www.nature.com/articles/s41467-021-26419-0},
volume = {12},
year = {2021}
}

@article{Diler2020,
archivePrefix = {arXiv},
arxivId = {1909.08778},
author = {Diler, Berk and Whiteley, Samuel J. and Anderson, Christopher P. and Wolfowicz, Gary and Wesson, Marie E. and Bielejec, Edward S. and {Joseph Heremans}, F. and Awschalom, David D.},
doi = {10.1038/s41534-020-0247-7},
eprint = {1909.08778},
issn = {2056-6387},
journal = {npj Quantum Information},
month = {Jan},
number = {1},
pages = {11},
publisher = {Springer US},
title = {{Coherent control and high-fidelity readout of chromium ions in commercial silicon carbide}},
url = {http://dx.doi.org/10.1038/s41534-020-0247-7 http://www.nature.com/articles/s41534-020-0247-7 https://www.nature.com/articles/s41534-020-0247-7},
volume = {6},
year = {2020}
}

@article{Harmon2022,
author = {Harmon, K J and Delegan, N and Highland, M J and He, H and Zapol, P and Heremans, F J and Hruszkewycz, S O},
doi = {10.1088/2633-4356/ac6b76},
issn = {2633-4356},
journal = {Materials for Quantum Technology},
month = {Jun},
number = {2},
pages = {023001},
title = {{Designing silicon carbide heterostructures for quantum information science: challenges and opportunities}},
url = {https://iopscience.iop.org/article/10.1088/2633-4356/ac6b76},
volume = {2},
year = {2022}
}

@article{Gottscholl2022,
archivePrefix = {arXiv},
arxivId = {2203.00329},
author = {Gottscholl, Andreas and Wagenh{\"{o}}fer, Maximilian and Klimmer, Manuel and Scherbel, Selina and Kasper, Christian and Baianov, Valentin and Astakhov, Georgy V. and Dyakonov, Vladimir and Sperlich, Andreas},
doi = {10.3389/fphot.2022.886354},
eprint = {2203.00329},
issn = {2673-6853},
journal = {Frontiers in Photonics},
month = {May},
number = {May},
pages = {1--9},
title = {{Superradiance of Spin Defects in Silicon Carbide for Maser Applications}},
url = {https://www.frontiersin.org/articles/10.3389/fphot.2022.886354/full},
volume = {3},
year = {2022}
}

@article{Fischer2018,
archivePrefix = {arXiv},
arxivId = {1709.00052},
author = {Fischer, M. and Sperlich, A. and Kraus, H. and Ohshima, T. and Astakhov, G. V. and Dyakonov, V.},
doi = {10.1103/PhysRevApplied.9.054006},
eprint = {1709.00052},
issn = {2331-7019},
journal = {Physical Review Applied},
month = {May},
number = {5},
pages = {054006},
publisher = {American Physical Society},
title = {{Highly Efficient Optical Pumping of Spin Defects in Silicon Carbide for Stimulated Microwave Emission}},
url = {https://doi.org/10.1103/PhysRevApplied.9.054006 https://link.aps.org/doi/10.1103/PhysRevApplied.9.054006},
volume = {9},
year = {2018}
}

@article{Castelletto2020,
author = {Castelletto, Stefania and Boretti, Alberto},
doi = {10.1088/2515-7647/ab77a2},
issn = {2515-7647},
journal = {Journal of Physics: Photonics},
month = {Apr},
number = {2},
pages = {022001},
publisher = {IOP Publishing},
title = {{Silicon carbide color centers for quantum applications}},
url = {https://iopscience.iop.org/article/10.1088/2515-7647/ab77a2},
volume = {2},
year = {2020}
}

@article{Castelletto2024,
author = {Castelletto, S. and Lew, C. T.K. and Lin, Wu-Xi and Xu, Jin-Shi},
doi = {10.1088/1361-6633/ad10b3},
issn = {0034-4885},
journal = {Reports on Progress in Physics},
month = {Jan},
number = {1},
pages = {014501},
pmid = {38029424},
title = {{Quantum systems in silicon carbide for sensing applications}},
url = {https://iopscience.iop.org/article/10.1088/1361-6633/ad10b3},
volume = {87},
year = {2024}
}

@article{Koehl2017,
author = {Koehl, William F. and Diler, Berk and Whiteley, Samuel J. and Bourassa, Alexandre and Son, N. T. and Janz{\'{e}}n, Erik and Awschalom, David D.},
doi = {10.1103/PhysRevB.95.035207},
issn = {2469-9950},
journal = {Physical Review B},
month = {Jan},
number = {3},
pages = {035207},
title = {{Resonant optical spectroscopy and coherent control of Cr$^{4+}$ spin ensembles in SiC and GaN}},
url = {https://link.aps.org/doi/10.1103/PhysRevB.95.035207},
volume = {95},
year = {2017}
}

@article{Falk2013,
author = {Falk, Abram L. and Buckley, Bob B. and Calusine, Greg and Koehl, William F. and Dobrovitski, Viatcheslav V. and Politi, Alberto and Zorman, Christian A. and Feng, Philip X.L. and Awschalom, David D.},
doi = {10.1038/ncomms2854},
issn = {2041-1723},
journal = {Nature Communications},
month = {May},
number = {1},
pages = {1819},
pmid = {23652007},
publisher = {Nature Publishing Group},
title = {{Polytype control of spin qubits in silicon carbide}},
url = {https://www.nature.com/articles/ncomms2854},
volume = {4},
year = {2013}
}

@article{Castelletto2022,
   author = {Stefania Castelletto and Alberto Peruzzo and Cristian Bonato and Brett C. Johnson and Marina Radulaski and Haiyan Ou and Florian Kaiser and Joerg Wrachtrup},
   doi = {10.1021/acsphotonics.1c01775},
   issn = {2330-4022},
   issue = {5},
   journal = {ACS Photonics},
   month = {5},
   pages = {1434-1457},
   publisher = {American Chemical Society},
   title = {Silicon Carbide Photonics Bridging Quantum Technology},
   volume = {9},
   url = {https://pubs.acs.org/doi/10.1021/acsphotonics.1c01775},
   year = {2022}
}

@article{Baur1997,
author = {Baur, J. and Kunzer, M. and Schneider, J.},
doi = {10.1002/1521-396X(199707)162:1<153::AID-PSSA153>3.0.CO;2-3},
issn = {00318965},
journal = {Physica status solidi (a)},
month = {Jul},
number = {1},
pages = {153--172},
title = {{Transition Metals in SiC Polytypes, as Studied by Magnetic Resonance Techniques}},
url = {https://onlinelibrary.wiley.com/doi/10.1002/1521-396X(199707)162:1%3C153::AID-PSSA153%3E3.0.CO;2-3},
volume = {162},
year = {1997}
}

@article{Sun2023b,
author = {Sun, Tianze and Xu, Zongwei and Wu, Jintong and Fan, Yexin and Ren, Fei and Song, Ying and Yang, Long and Tan, Pingheng},
doi = {10.1016/j.ceramint.2022.10.219},
issn = {02728842},
journal = {Ceramics International},
month = {Mar},
number = {5},
pages = {7452--7465},
publisher = {Elsevier Ltd},
title = {{Divacancy and silicon vacancy color centers in 4H-SiC fabricated by hydrogen and dual ions implantation and annealing}},
url = {https://doi.org/10.1016/j.ceramint.2022.10.219 https://linkinghub.elsevier.com/retrieve/pii/S0272884222038007},
volume = {49},
year = {2023}
}

@article{Nagy2019,
archivePrefix = {arXiv},
arxivId = {1810.10296},
author = {Nagy, Roland and Niethammer, Matthias and Widmann, Matthias and Chen, Yu-Chen and Udvarhelyi, P{\'{e}}ter and Bonato, Cristian and Hassan, Jawad Ul and Karhu, Robin and Ivanov, Ivan G. and Son, Nguyen Tien and Maze, Jeronimo R. and Ohshima, Takeshi and Soykal, {\"{O}}ney O. and Gali, {\'{A}}d{\'{a}}m and Lee, Sang-Yun and Kaiser, Florian and Wrachtrup, J{\"{o}}rg},
doi = {10.1038/s41467-019-09873-9},
eprint = {1810.10296},
issn = {2041-1723},
journal = {Nature Communications},
month = {Apr},
number = {1},
pages = {1954},
pmid = {31028260},
publisher = {Springer US},
title = {{High-fidelity spin and optical control of single silicon-vacancy centres in silicon carbide}},
url = {http://dx.doi.org/10.1038/s41467-019-09873-9 https://www.nature.com/articles/s41467-019-09873-9},
volume = {10},
year = {2019}
}

@article{Bottger2016b,
author = {B{\"{o}}ttger, Thomas and Thiel, C. W. and Cone, R. L. and Sun, Y. and Faraon, A.},
doi = {10.1103/PhysRevB.94.045134},
issn = {2469-9950},
journal = {Physical Review B},
month = {Jul},
number = {4},
pages = {045134},
title = {{Optical spectroscopy and decoherence studies of Yb$^{3+}$:YAG at 968 nm}},
url = {https://link.aps.org/doi/10.1103/PhysRevB.94.045134},
volume = {94},
year = {2016}
}

@article{Babin2022,
author = {Babin, Charles and St{\"{o}}hr, Rainer and Morioka, Naoya and Linkewitz, Tobias and Steidl, Timo and W{\"{o}}rnle, Raphael and Liu, Di and Hesselmeier, Erik and Vorobyov, Vadim and Denisenko, Andrej and Hentschel, Mario and Gobert, Christian and Berwian, Patrick and Astakhov, Georgy V. and Knolle, Wolfgang and Majety, Sridhar and Saha, Pranta and Radulaski, Marina and Son, Nguyen Tien and Ul-Hassan, Jawad and Kaiser, Florian and Wrachtrup, J{\"{o}}rg},
doi = {10.1038/s41563-021-01148-3},
issn = {1476-1122},
journal = {Nature Materials},
month = {Jan},
number = {1},
pages = {67--73},
pmid = {34795400},
publisher = {Springer US},
title = {{Fabrication and nanophotonic waveguide integration of silicon carbide colour centres with preserved spin-optical coherence}},
url = {https://www.nature.com/articles/s41563-021-01148-3},
volume = {21},
year = {2022}
}

@article{Yan2020,
archivePrefix = {arXiv},
arxivId = {2004.06261},
author = {Yan, Fei-Fei and Yi, Ai-Lun and Wang, Jun-Feng and Li, Qiang and Yu, Pei and Zhang, Jia-Xiang and Gali, Adam and Wang, Ya and Xu, Jin-Shi and Ou, Xin and Li, Chuan-Feng and Guo, Guang-Can},
doi = {10.1038/s41534-020-0270-8},
eprint = {2004.06261},
issn = {2056-6387},
journal = {npj Quantum Information},
month = {May},
number = {1},
pages = {38},
publisher = {Springer US},
title = {{Room-temperature coherent control of implanted defect spins in silicon carbide}},
url = {http://dx.doi.org/10.1038/s41534-020-0270-8 https://www.nature.com/articles/s41534-020-0270-8},
volume = {6},
year = {2020}
}

@article{Seo2016,
author = {Seo, Hosung and Falk, Abram L. and Klimov, Paul V. and Miao, Kevin C. and Galli, Giulia and Awschalom, David D.},
doi = {10.1038/ncomms12935},
issn = {2041-1723},
journal = {Nature Communications},
month = {Sep},
number = {1},
pages = {12935},
publisher = {Nature Publishing Group},
title = {{Quantum decoherence dynamics of divacancy spins in silicon carbide}},
url = {http://dx.doi.org/10.1038/ncomms12935 https://www.nature.com/articles/ncomms12935},
volume = {7},
year = {2016}
}

@article{Klimov2014,
archivePrefix = {arXiv},
arxivId = {1310.4844},
author = {Klimov, P. V. and Falk, A. L. and Buckley, B. B. and Awschalom, D. D.},
doi = {10.1103/PhysRevLett.112.087601},
eprint = {1310.4844},
issn = {0031-9007},
journal = {Physical Review Letters},
month = {Feb},
number = {8},
pages = {087601},
title = {{Electrically Driven Spin Resonance in Silicon Carbide Color Centers}},
url = {https://link.aps.org/doi/10.1103/PhysRevLett.112.087601},
volume = {112},
year = {2014}
}

@article{Niethammer2019,
author = {Niethammer, Matthias and Widmann, Matthias and Rendler, Torsten and Morioka, Naoya and Chen, Yu-Chen and St{\"{o}}hr, Rainer and Hassan, Jawad Ul and Onoda, Shinobu and Ohshima, Takeshi and Lee, Sang-Yun and Mukherjee, Amlan and Isoya, Junichi and Son, Nguyen Tien and Wrachtrup, J{\"{o}}rg},
doi = {10.1038/s41467-019-13545-z},
issn = {2041-1723},
journal = {Nature Communications},
month = {Dec},
number = {1},
pages = {5569},
pmid = {31804489},
publisher = {Springer US},
title = {{Coherent electrical readout of defect spins in silicon carbide by photo-ionization at ambient conditions}},
url = {http://dx.doi.org/10.1038/s41467-019-13545-z https://www.nature.com/articles/s41467-019-13545-z},
volume = {10},
year = {2019}
}

@article{Lekavicius2022,
author = {Lekavicius, I. and Myers-Ward, R.L. and Pennachio, D.J. and Hajzus, J.R. and Gaskill, D.K. and Purdy, A.P. and Yeats, A.L. and Brereton, P.G. and Glaser, E.R. and Reinecke, T.L. and Carter, S.G.},
doi = {10.1103/PRXQuantum.3.010343},
issn = {2691-3399},
journal = {PRX Quantum},
month = {Mar},
number = {1},
pages = {010343},
publisher = {American Physical Society},
title = {{Orders of Magnitude Improvement in Coherence of Silicon-Vacancy Ensembles in Isotopically Purified 4H-SiC}},
url = {https://doi.org/10.1103/PRXQuantum.3.010343 https://link.aps.org/doi/10.1103/PRXQuantum.3.010343},
volume = {3},
year = {2022}
}

@article{Gilardoni2020,
archivePrefix = {arXiv},
arxivId = {1912.04612},
author = {Gilardoni, Carmem M. and Bosma, Tom and van Hien, Danny and Hendriks, Freddie and Magnusson, Bj{\"{o}}rn and Ellison, Alexandre and Ivanov, Ivan G. and Son, N. T. and van der Wal, Caspar H},
doi = {10.1088/1367-2630/abbf23},
eprint = {1912.04612},
issn = {1367-2630},
journal = {New Journal of Physics},
month = {Oct},
number = {10},
pages = {103051},
publisher = {IOP Publishing},
title = {{Spin-relaxation times exceeding seconds for color centers with strong spin–orbit coupling in SiC}},
url = {https://iopscience.iop.org/article/10.1088/1367-2630/abbf23},
volume = {22},
year = {2020}
}

@article{Wolfowicz2020,
archivePrefix = {arXiv},
arxivId = {1908.09817},
author = {Wolfowicz, Gary and Anderson, Christopher P. and Diler, Berk and Poluektov, Oleg G. and Heremans, F. Joseph and Awschalom, David D.},
doi = {10.1126/sciadv.aaz1192},
eprint = {1908.09817},
issn = {2375-2548},
journal = {Science Advances},
month = {May},
number = {18},
pages = {2--9},
pmid = {32426475},
title = {{Vanadium spin qubits as telecom quantum emitters in silicon carbide}},
url = {https://www.science.org/doi/10.1126/sciadv.aaz1192},
volume = {6},
year = {2020}
}

@article{Singh2024b,
   author = {Harpreet Singh and Noella D’Souza and Joseph Garrett and Angad Singh and Brian Blankenship and Emanuel Druga and Riccardo Montis and Liang Z. Tan and Ashok Ajoy},
   doi = {10.1038/s41467-025-65508-2},
   issn = {2041-1723},
   issue = {1},
   journal = {Nature Communications},
   month = {11},
   pages = {10530},
   title = {High sensitivity pressure and temperature quantum sensing in pentacene-doped p-terphenyl single crystals},
   volume = {16},
   url = {https://www.nature.com/articles/s41467-025-65508-2},
   year = {2025}
}

@article{Sukachev2017,
archivePrefix = {arXiv},
arxivId = {1708.08852},
author = {Sukachev, D. D. and Sipahigil, Alp and Nguyen, C. T. and Bhaskar, M. K. and Evans, R. E. and Jelezko, Fedor and Lukin, M. D.},
doi = {10.1103/PhysRevLett.119.223602},
eprint = {1708.08852},
issn = {0031-9007},
journal = {Physical Review Letters},
month = {Nov},
number = {22},
pages = {223602},
pmid = {29286819},
title = {{Silicon-Vacancy Spin Qubit in Diamond: A Quantum Memory Exceeding 10 ms with Single-Shot State Readout}},
url = {http://arxiv.org/abs/1708.08852 http://dx.doi.org/10.1103/PhysRevLett.119.223602 https://link.aps.org/doi/10.1103/PhysRevLett.119.223602},
volume = {119},
year = {2017}
}

@article{Giest1964,
author = {Giest, D. and {G. Romelt}},
journal = {Solid State Communications},
pages = {149},
title = {{Paramagnetische elektronenresonanz in bornitrid}},
volume = {2},
year = {1964}
}

\end{document}